\documentclass{aa}  
\usepackage[switch]{lineno}
\usepackage{amsmath,amstext}
\usepackage[T1]{fontenc}
\usepackage{xcolor}
\usepackage{soul}
\usepackage{ulem}
\usepackage{etoolbox}
\robustify{\sout}
\usepackage{upgreek}
\usepackage{hyperref}
\usepackage{float}
\usepackage{orcidlink}
\usepackage{multirow}
\usepackage[utf8]{inputenc}
\usepackage{natbib}
\usepackage{booktabs}
\usepackage{longtable}
\usepackage[flushleft]{threeparttable}
\usepackage{tabularx}
\usepackage{makecell}
\usepackage[fleqn]{amsmath}
\usepackage{hyperref}

\usepackage{wrapfig} 
\hypersetup{
  colorlinks   = true,    
  urlcolor     = blue,    
  linkcolor    = blue,   
  citecolor    = blue      
}

\usepackage{graphicx}

\usepackage{txfonts}

\newcommand{\Msun}{\mathrm{M}_\odot}

\begin{document}

\authorrunning{Meng,~L. et al.}
\titlerunning{The double neutron star PSR~J1946$+$2052}

   \title{The double neutron star PSR~J1946$+$2052}
   \subtitle{I. Masses and tests of general relativity}

   \author{Lingqi~Meng\inst{1,2,3}\orcidlink{0000-0002-2885-568X},
          Paulo~C.~C.~Freire\inst{2}\orcidlink{0000-0003-1307-9435},
          Kevin~Stovall\inst{4}\orcidlink{https://orcid.org/0000-0002-7261-594X},
          Norbert~Wex\inst{2}\orcidlink{https://orcid.org/0000-0003-4058-2837}
          \and
          Xueli~Miao\inst{1,2}\orcidlink{https://orcid.org/0000-0003-1185-8937}
          \and
          Weiwei~Zhu\inst{1,5}\orcidlink{https://orcid.org/0000-0001-5105-4058}
          \and
          Michael~Kramer\inst{2,6}\orcidlink{https://orcid.org/0000-0002-4175-2271}
          \and
          James~M.~Cordes\inst{7}\orcidlink{https://orcid.org/0000-0002-4049-1882}
          \and
          Huanchen~Hu\inst{2}\orcidlink{https://orcid.org/0000-0001-5105-4058}
          \and
          Jinchen~Jiang\inst{2}\orcidlink{https://orcid.org/0000-0002-6465-0091}
          Emilie~Parent\inst{8,9}\orcidlink{https://orcid.org/0000-0002-0430-6504}
          \and
          Lijing~Shao\inst{10,1,2}\orcidlink{https://orcid.org/0000-0002-1334-8853}
          \and
          Ingrid~H.~Stairs\inst{11}\orcidlink{https://orcid.org/0000-0001-9784-8670}
          \and
          Mengyao~Xue\inst{1}\orcidlink{https://orcid.org/0000-0001-8018-1830}
          \and
          Adam~Brazier\inst{12}\orcidlink{https://orcid.org/0000-0001-6341-7178}
          \and
          Fernando~Camilo\inst{13}\orcidlink{https://orcid.org/0000-0002-1873-3718}
          \and
          David~J.~Champion\inst{2}\orcidlink{https://orcid.org/0000-0003-1361-7723}
          \and
          Shami~Chatterjee\inst{7}\orcidlink{ https://orcid.org/0000-0002-2878-1502}
          \and
          Fronefield~Crawford\inst{14}\orcidlink{https://orcid.org/0000-0002-2578-0360}
          \and
          Ziyao~Fang\inst{1}
          \and
          Qiuyang~Fu\inst{1,3}
          \and
          Yanjun~Guo\inst{1}
          \and
          Jason~W.~T.~Hessels\inst{15,16}\orcidlink{https://orcid.org/0000-0003-2317-1446}
          \and
          Maura MacLaughlin\inst{17}\orcidlink{https://orcid.org/0000-0001-7697-7422}
          \and
          Chenchen~Miao\inst{18}\orcidlink{https://orcid.org/0000-0002-9441-2190}
          \and
          Jiarui~Niu\inst{1}\orcidlink{https://orcid.org/0000-0001-8065-4191}
          \and
          Ziwei~Wu\inst{1}\orcidlink{https://orcid.org/0000-0002-1381-7859}
          \and
          Jumei~Yao\inst{19}\orcidlink{https://orcid.org/0000-0002-4997-045X}
          \and
          Mao~Yuan\inst{20}\orcidlink{https://orcid.org/0000-0003-1874-0800}
          \and
          Youlin~Yue\inst{1}\orcidlink{https://orcid.org/0000-0003-4415-2148}
          \and
          Chengmin~Zhang\inst{1}
          }

   \institute{National Astronomical Observatories, Chinese Academy of Sciences, Beijing 100101, China\\
    \email{menglingqi@nao.cas.cn, zhuww@nao.cas.cn}
   \and 
   Max-Planck-Institut f\"{u}r Radioastronomie, Auf dem H\"{u}gel 69, 53131 Bonn, Germany
    \and
    School of Astronomy and Space Science, University of Chinese Academy of Sciences, Beijing 100049, China
   \and
    University of New Mexico, Department of Physics and Astronomy, University of New Mexico, 210 Yale Blvd NE, Albuquerque, NM 87106, USA
    \and
    Institute for Frontiers in Astronomy and Astrophysics, Beijing Normal University, Beijing 102206, China
    \and
    Jodrell Bank Centre for Astrophysics, Department of Physics and Astronomy, University of Manchester, M13 9PL, Manchester, UK
    \and
    Cornell Centre for Astrophysics and Planetary Science and Department of Astronomy, Cornell University, Ithaca, NY 14853, USA
    \and
    Institute of Space Sciences (ICE, CSIC), Campus UAB, Carrer de Can Magrans s/n, 08193, Barcelona, Spain
    \and
    Institut d'Estudis Espacials de Catalunya (IEEC), Carrer Gran Capit\`a 2--4, 08034 Barcelona, Spain
    \and
    Kavli Institute for Astronomy and Astrophysics, Peking University, Beijing 100871, China
    \and
    Department of Physics and Astronomy, University of British Columbia, 6224 Agricultural Rd., Vancouver, BC V6T 1Z1, Canada
    \and
    Cornell Center for Advanced Computing, Cornell University, Ithaca, NY 14853, USA
    \and
    South African Radio Astronomy Observatory, Liesbeek House, River Park, Cape Town 7700, South Africa
    \and
    Department of Physics and Astronomy, Franklin and Marshall College, P.O. Box 3003, Lancaster, PA 17604, USA
    \and
ASTRON, the Netherlands Institute for Radio Astronomy, Oude Hoogeveensedijk 4, 7991 PD Dwingeloo, The Netherlands
    \and
Anton Pannekoek Institute for Astronomy, University of Amsterdam, Science Park 904, 1098 XH, Amsterdam, The Netherlands
    \and
     Department of Physics and Astronomy and Center for Gravitational Waves and Cosmology, West Virginia University, Morgantown, WV 26506-6315
    \and 
    Research Institute of Artificial Intelligence, Zhejiang Lab, Hangzhou, Zhejiang 311121, China
    \and
    Xinjiang Astronomical Observatory, Chinese Academy of Sciences, 150, Science 1-Street, Urumqi, Xinjiang 830011, China
    \and
    National Space Science Center, Chinese Academy of Sciences, Beijing 100190, China
             }

   \date{Received September 15, 1996; accepted March 16, 1997}

  \abstract

   {Double neutron star (DNS) systems are superb laboratories for testing theories of gravity and constraining the equation of state of ultra-dense matter. PSR~J1946+2052 is a particularly intriguing DNS system due to its orbital period (1h 53m), the shortest among all DNS systems known in our Galaxy.}
   {We aim to conduct high-precision timing of PSR J1946+2052 to determine the masses of the two neutron stars in the system, test general relativity (GR) and assess the system’s potential for future measurement of the moment of inertia of the pulsar.}
   {We analysed seven years of timing data from the Arecibo 305-m radio telescope, the Green Bank Telescope (GBT), and the Five-hundred-meter Aperture Spherical radio Telescope (FAST). The data processing accounted for dispersion measure (DM) variations and relativistic spin precession-induced profile evolution. We employed both theory-independent (DDFWHE) and GR-dependent (DDGR) binary models to measure the spin parameters, kinematic parameters and orbital parameters.}
   {The timing campaign has resulted in the precise measurement of five post-Keplerian parameters, which yield very precise masses for the system  (total mass $M = 2.531858(60) \, \rm M_{\odot}$, companion mass $M_{\rm c} = 1.2480(21) \, \rm M_{\odot}$ and pulsar mass $M_{\rm p} = 1.2838(21) \, \rm M_{\odot}$) and three tests of general relativity. 
One of these is the second most precise test of the radiative properties of gravity to date: the intrinsic orbital decay, $\dot{P}_{\rm b,int} = -1.8288(16) \times 10^{-12} \rm \, s\, s^{-1}$, represents
$1.00005(91)$ of the GR prediction, indicating that the theory has passed this stringent test. The other two tests, of the Shapiro delay parameters, have precisions of 6\% and 5\% respectively; this is caused by the moderate orbital inclination of the system, $\sim 74^{\circ}$; the
measurements of the Shapiro delay parameters also agree with the GR predictions.
Additionally, we analyse the higher-order contributions of $\dot{\omega}$, including the Lense-Thirring contribution.
Both the second post-Newtonian and the Lense-Thirring contributions are larger than the current uncertainty of $\dot{\omega}$ ($\delta\dot{\omega}=4\times10^{-4}\,\rm deg\,yr^{-1}$), leading to the higher-order correction for the total mass.

}
   {}
   {}

    \keywords{pulsars: individual (PSR J1946+2052) $-$ gravitation $-$ binary: close}
  
   \maketitle

\section{Introduction}
\label{sec:intro}

The discovery of the first binary pulsar, PSR~B1913+16 \citep{1975ApJ...195L..51H} opened the era of gravitational wave (GW) astronomy. Indeed, the detailed timing of the pulsar enabled the detection of three relativistic effects in its orbital motion: 1) an increase in the longitude of periastron with time (parametrised by its rate, $\dot{\omega}$), 2) the combination of the varying relativistic time dilation and gravitational redshift with orbital phase (parametrised by a physical quantity known as the amplitude of the Einstein delay, $\gamma$) and 3) a decrease of the orbital period with time \citep[again, parametrised by its rate, $\dot{P}_{\rm b}$,][]{1982ApJ...253..908T,1989ApJ...345..434T}. These ``Post-Keplerian'' (PK) parameters quantify relativistic effects in the orbital motion and in the propagation of light for all fully conservative, boost-invariant gravity theories \citep{1992PhRvD..45.1840D}. In general relativity (GR), they depend only on the two masses of the components of the system, at least to the leading order. Thus, if we are sure the measured orbital effects are relativistic\footnote{Using arguments based on binary stellar evolution (for a review, see \citealt{2017ApJ...846..170T}), it is very likely that the companion to PSR B1913+16 is another neutron star (NS), which remains to this day undetected; this implies that the system is ``clean'', i.e., the orbital motion is that of two point masses in free fall around each other. This is important to establish that the observed effects in the timing are indeed relativistic.} and assume GR as the correct theory of gravity, then with two measured PK parameters, we can determine the two masses in the binary.
Measuring a third PK effect - the  $\dot{P}_{\rm b}$ - provided a test of the self-consistency of the theory. And, indeed, the observed  $\dot{P}_{\rm b}$ matched the GR prediction for the orbital decay of the system caused by GW damping.
This was the first test of the radiative properties of gravity, and the first test of the gravitational properties of strongly self-gravitating objects; these two aspects of gravity cannot be studied in the laboratory of the Solar System \citep[for a review, see][]{2024LRR....27....5F}.

The observation of the orbital decay in this system represented the first evidence of gravitational waves. This was especially important as the reality of GWs had not yet been fully established theoretically. 
Its many astrophysical implications (like the inevitability of double neutron star mergers) paved the path for the development of ground-based GW detectors. 

In the following decades, a few hundred binary radio pulsars have been discovered, but of these, only 24 are confirmed as double neutron star systems (DNSs)\footnote{See \url{https://www3.mpifr-bonn.mpg.de/staff/pfreire/NS_masses.html\#part3}}. Of special importance for the discussion in this paper is the ``double pulsar'' system, PSR~J0737$-$3039A/B. With an orbital period of 2h 27 m, it was at the time of its discovery the DNS with the shortest coalescence time known, about 86 Myr, a fact that significantly increased the expected rate of such mergers \citep{2003Natur.426..531B}. 
Apart from that, this is the only DNS known where both NSs are radio pulsars \citep{2004Sci...303.1153L}; furthermore, its orbital inclination is the highest known for any binary pulsar, yielding extremely precise measurements of the Shapiro delay.

Continued radio timing of this system has yielded not only the most precise NS masses, but a total of several independent tests of GR, all of which the theory passes \citep[Table V in][]{2021PhRvX..11d1050K}. These include the most precise test of the quadrupole formula for GW damping, which is 25 times more precise than the second best system (the aforementioned test with the Hulse-Taylor pulsar, \citealt{2016ApJ...829...55W}); the test of the propagation of light in a spacetime with a curvature that is $10^3$ times larger than near Sagittarius A* and the detection, for the first time, of next-to-leading order effects (see the improvement of the test in the signal propagation with the MeerKat data in \citealt{2022A&A...667A.149H}). The system has also allowed, for the first time, the derivation of a useful upper limit of the moment of inertia ($I < 3 \times 10^{45} \, \rm g \, cm^2$; 90\% C. L.) of a radio pulsar, PSR~J0737$-$3039A.
The results from this system are directly relevant to this work and will be discussed in more detail below.

The focus of this work, PSR J1946$+$2052, was discovered in 2017 by the
Arecibo L-Band Feed Array pulsar (PALFA) survey~\citep{2006ApJ...637..446C,2015ApJ...812...81L}
and reported by \cite{2018ApJ...854L..22S}, henceforth Paper~2018.
Of all known Galactic DNSs, PSR~J1946+2052 is the one with the shortest orbital period, 1h 53 m, and the shortest coalescence time, 46 Myr. The pulsar's spin period is the smallest for any known member of a DNS, $\sim$17 ms; the known spin-down means that the pulsar has the largest spin parameter at merger among all Galactic DNSs ( Paper~2018). Its orbital eccentricity ($e \, = \, 0.063848(9)$) is the smallest for any Galactic DNS, even lower than that of PSR~J1325$-$6253 ($e = 0.0640091(7)$, \citealt{2022MNRAS.512.5782S}).
Interestingly, the orbital period and eccentricity of PSR~J1946+2052 are very similar to those the Double Pulsar system will have in about 40 Myr, almost halfway through its merger.
Given the characteristic age of PSR~J1946+2052, the upper limits in the orbital eccentricity and period at birth are $e\textless0.14$ and $P_{\rm b}\textless0.17$ days \citep{2018ApJ...854L..22S}, which makes it very likely that they have decreased significantly since the formation of the system owing to GW emission.

The published results on this system were based on a very small timing baseline of 71 days. Despite that short baseline, a remarkably accurate measurement of the spin period and spin period derivative could be established, thanks to the precise position obtained with the Very Large Array. The spin parameters imply a characteristic age of about 290 Myr and a surface magnetic field of about $4 \, \times \, 10^9 \, \rm G$. Also remarkable was by far the largest periastron advance for any known DNS system in the Galaxy, $\dot{\omega} = 25.6 \pm 0.3 \rm \, \deg \, yr^{-1}$; this implies (assuming the validity of GR) a total system mass of $2.50 \pm 0.04 \, \mathrm{M_{\odot}}$, making this one of the least massive DNSs known (see also \citealt{2017ApJ...851L..29M}).

Since the publication of Paper~2018, we have carried out regular radio timing observations of this system by using the Arecibo 305-m radio telescope until the end of its operational time in 2020. Since 2019, we have also been observing the system with the Five-hundred-meter Aperture Spherical radio Telescope \citep[FAST, ][]{jiang2020fundamental}. Based on the FAST data,
\cite{2024ApJ...966...46M} reported the significant profile evolution of PSR~J1946+2052, which indicates that its spin axis is misaligned with the orbital angular momentum and is undergoing relativistic spin precession. The first report of the relativistic spin precession was in PSR~B1913+16, where \cite{weisberg1989evidence} and \cite{kramer1998determination} observed a secular variation of the relative amplitude of the two prominent components of PSR B1913+16’s profile. Such pulse evolution will introduce time offsets while using one standard template to generate times-of-arrival (ToAs), which should be compensated correctly.

This paper discusses some of the results of these radio observations, especially the timing. Its structure is as follows:
In Sect.~\ref{sec:observations}, we describe the observations themselves, the data processing and ToA generation and analysis. 
In Sect.~\ref{sec:results}, we present the new timing results, in particular the proper motion of the system and PK parameters.
In Sect.~\ref{sec:masses}, we discuss the masses of the two neutron stars in this system and the three GR tests that are now possible with this system.
The higher-order contributions in $\dot{\omega}$ are also considered, leading to the correction to the total mass.
We finally summarise our results in Sect.~\ref{sec:conclusions}.

\section{Observations and data analysis}
\label{sec:observations}

\begin{figure}
\begin{center}
\includegraphics[width=\columnwidth]{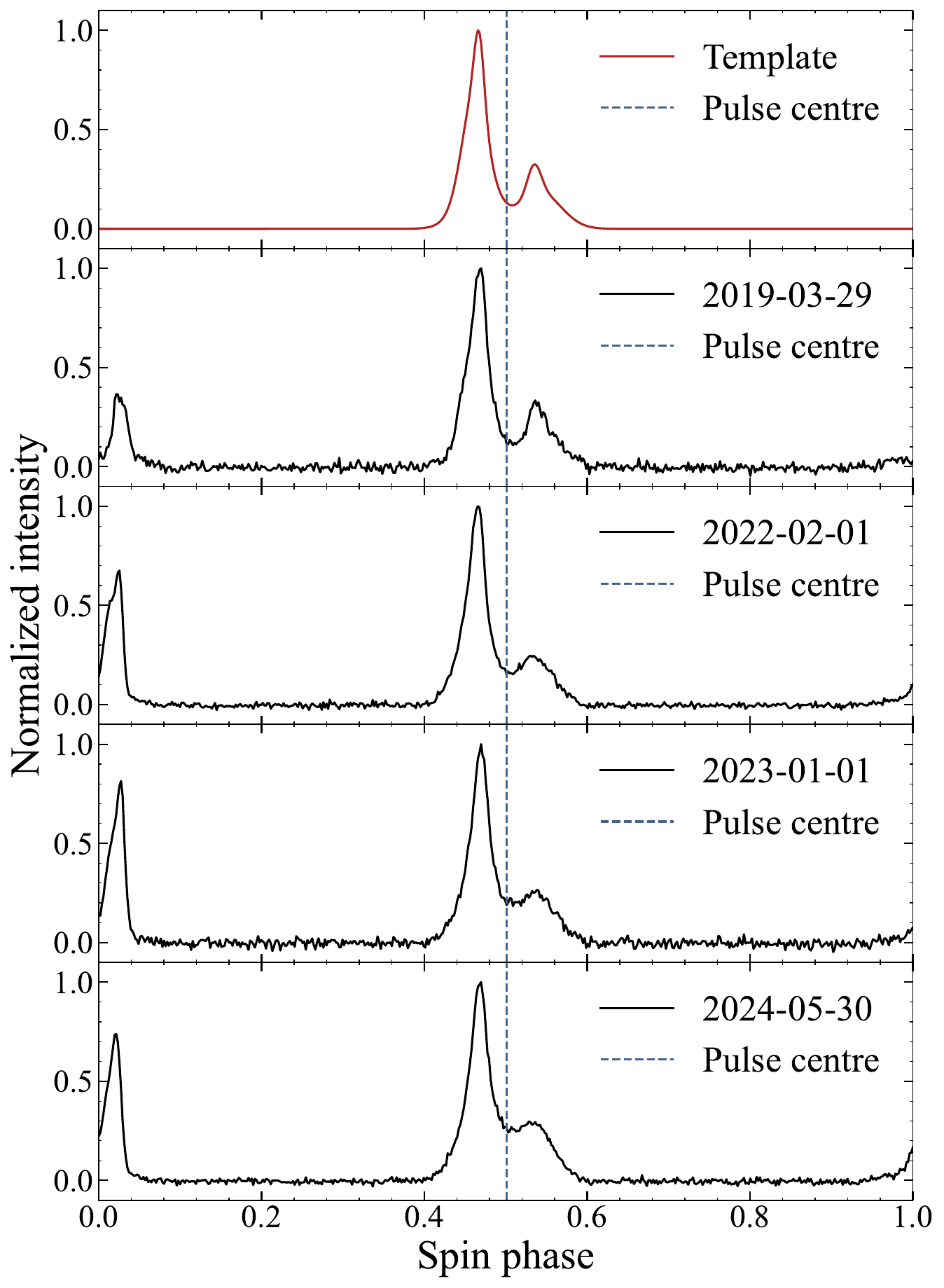}
\caption{The frequency-averaged template we used to fit ToAs and some examples of the observed pulse profile (Stokes I) are displayed in red and black solid lines, respectively. We generated the template with the pulse profile of 2019-03-29 and aligned each profile with the centre of the two Gaussian functions that we used to fit the pulse, indicated by the blue dashed lines. The less severe profile evolution in the main pulse can be seen in this figure compared to that in the interpulse, which makes it reasonable to fit ToAs only with the main pulse. One can also notice that the separation between the main pulse and the interpulse is increasing over time.}
\label{fig:profile}
\end{center}
\end{figure}

\begin{figure}
\begin{center}
\includegraphics[width=\columnwidth]{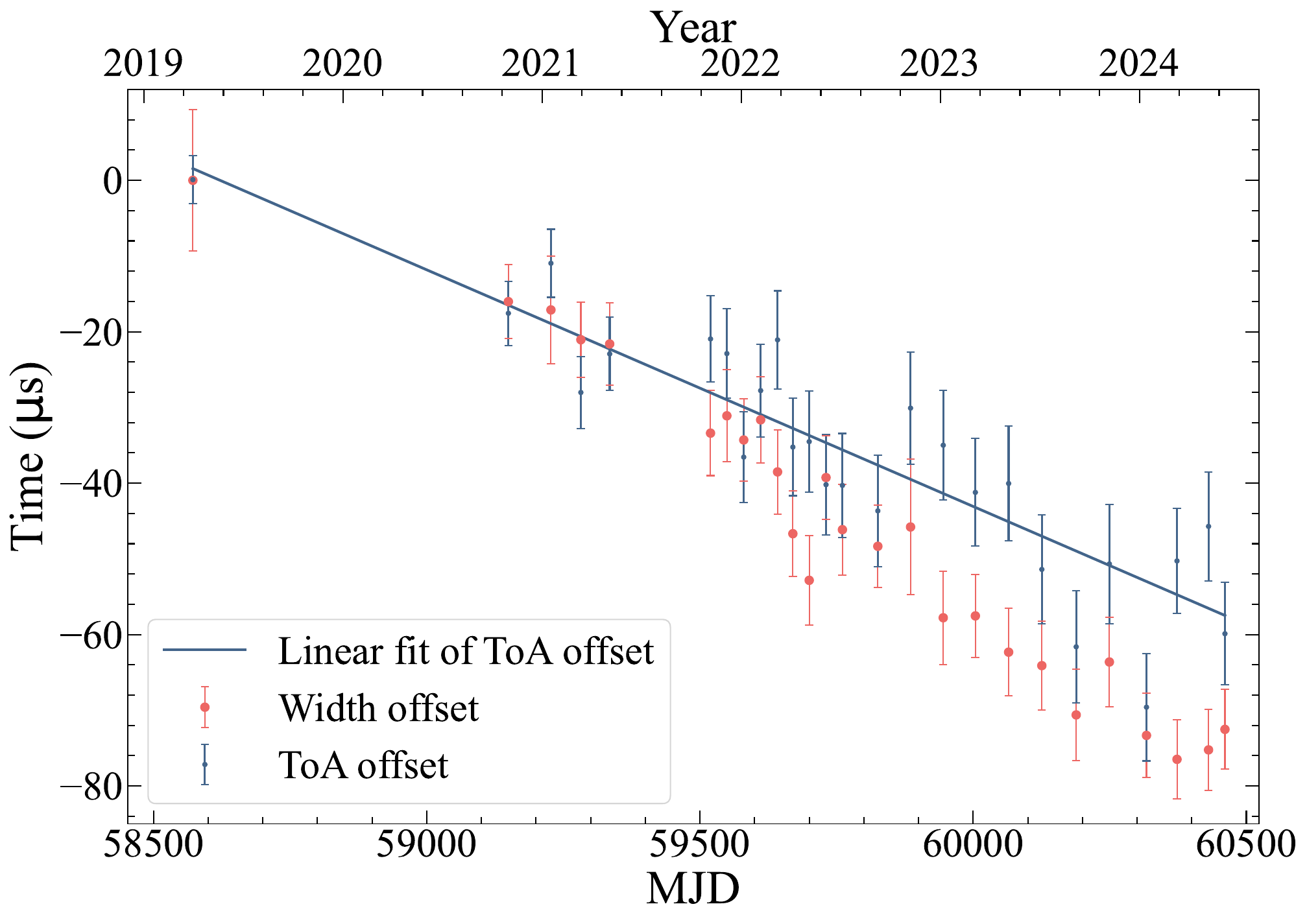}
\caption{ToA offsets derived from the standard template of the first observation and the integrated pulse profile are displayed in blue points. The pulse width (the unit is the same as ToA offset) of each integrated pulse profile is plotted in red points. The correlation between the ToA offsets and pulse widths indicates the strong influence introduced by the profile evolution on measuring the ToAs.
The blue solid line is the linear fit between MJD and ToA offsets, which is used to correct ToAs.}

\label{fig:offset}
\end{center}
\end{figure}

Most Arecibo observations in this work used the L-band receiver; for details, see Paper~2018. This set of observations ended on April 4, 2020. In addition, we have made a set of six observations with the Arecibo P-band receiver, which has a centre frequency of 327 MHz. All of these used the 
Puerto Rican Ultimate Pulsar Processing Instrument (PUPPI) back-end, which was based on its Green Bank predecessor (GUPPI, \citealt{2008SPIE.7019E..1DD}). We have also obtained one full-orbit observation with the 800 MHz receiver of the Green Bank Telescope, with GUPPI as a back-end.

The FAST observations were carried out with the centre beam of FAST's 19-beam receiver at a central frequency of 1250 MHz with a bandwidth of 500 MHz. The frequency resolution is 0.122\,MHz. 
The signal is digitised in 8-bit, converted into 4-polarisations with a sample time of 49.152$\,\upmu$s and de-dispersed incoherently.
Most of the FAST observations are 2 hours long. This yields high signal-to-noise ratio pulse profiles and, by covering full orbits, results in precise measurements of the orbital parameters. 
In Fig.~\ref{fig:profile}, we displayed several examples of the observed pulse profiles from FAST observations and the template we used to derive ToAs.

The initial timing analysis for PSR J1946$+$2052 was described in Paper~2018. As in that analysis, we generated multi-frequency ToAs and used the {\sc tempo} timing software\footnote{\url{http://tempo.sourceforge.net/}} to analyse them. First, the ToAs measured at the telescope are converted to the Bureau International des Poids et Mesures (BIPM) 2023 timescale. To subtract the telescope's motion relative to the Solar System barycenter, we have used the International Earth Rotation Service data and the Jet Propulsion Laboratory's DE 440 Solar System ephemeris \citep{Park_2021}. 

As reported in \cite{2024ApJ...966...46M}, in the FAST observations, the pulse profile of PSR~J1946+2052 is seen to change with time due to the relativistic spin precession, while in Arecibo observations, there's no sign of profile evolution. 
Dealing with the time offsets introduced by profile evolution is necessary when using a single template to generate ToAs. The same situation can be found in several other relativistic binaries, such as, e.g., PSR~B1913+16 \citep{2016ApJ...829...55W}, PSR B1534+12 \citep{2014ApJ...787...82F} and PSR~J1906+0746 \citep{Desvignes_2019}.
Unlike PSR~B1913+16, but similarly to what happens in PSR~J1906+0746, the profile of PSR~J1946+2052 has a main pulse and an interpulse. 
The separation between these two main components is increasing, as shown in Fig.~\ref{fig:profile}, and it is hard to identify the absolute movement of each pulse.
We chose to make a standard template with only the main pulse, owing to its greater observed stability.

\begin{figure}
    \centering
    \includegraphics[width=1.\linewidth]{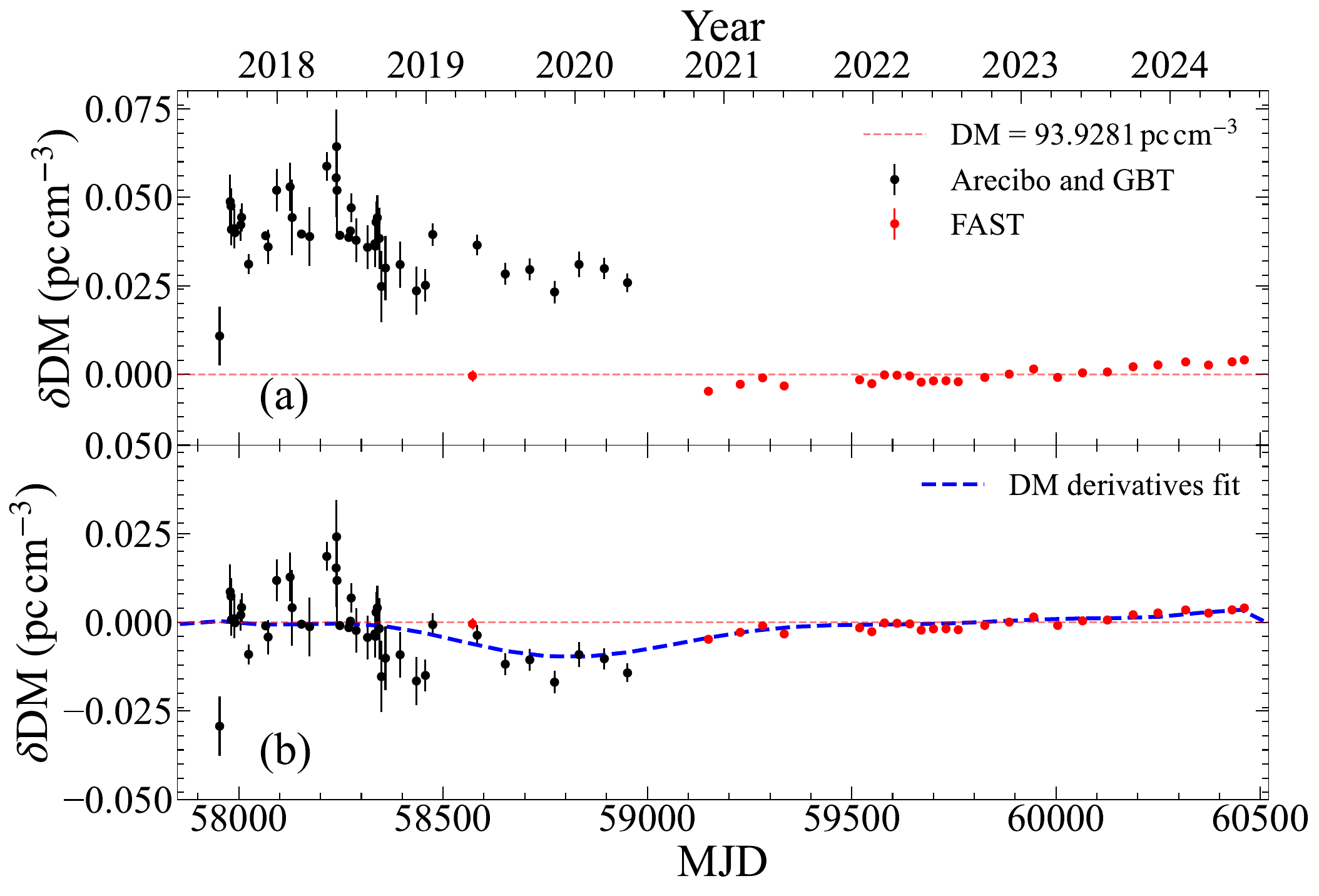}
    \caption{DM variations, derived from the DMX model with a time bin of 1 day, are displayed in this figure. Panels (a) and (b) represent the DM variation before and after the DM correction, respectively. DM measurements from Arecibo and GBT are represented by black points, and those from FAST are represented by red points. The red dashed line indicates the final measurement of the DM, which is 93.9281$\,\rm pc\, cm^{-3}$. We display the 10-order DM derivative fit in panel (b) with the blue dashed line.}
    \label{fig:DM_variation}
\end{figure}

In order to measure the time offsets caused by profile evolution, we introduced three assumptions: 1) the shape of the emission beam is a circular hollow cone \citep{1983ApJ...274..333R,kramer1998determination}, 2) the centre of the main pulse is stable in the spin phase because it is the region with the fastest change in the position angle of the linear polarisation, which according to the rotating vector model \citep{radhakrishnan1969magnetic} makes it the point in the rotation when one of the magnetic poles is closest to our line of sight; furthermore this region showed less significant evolution in \cite{2024ApJ...966...46M} 
and 3) we made a simplifying assumption, which is that, relative to this fiducial point, any time offsets in different frequency channels of each epoch are caused by changes in the DM only, not due to the pulse profile evolution with time\footnote{Generally, the use of 2-D templates implies that there are no TOA offsets with frequency caused by the profile change with frequency; however this does not guarantee that the same will happen after the pulse profile changes in time due to geodetic precession.}.

Firstly, we scrunched the time, polarisation and frequency information to get the high signal-to-noise ratio pulse profile of each observation.
We chose the first FAST observation (2019-03-29) as the template after making the template frequency-resolved by dividing it into 4 sub-bands and smoothing the pulse profile with \texttt{paas} in \texttt{PSRCHIVE} \citep{hotan2004psrchive}.
The frequency-averaged template is displayed in the top panel of Fig.~\ref{fig:profile}.
Secondly, we used two Gaussian functions to fit every main pulse and aligned each main pulse with the centre of the two Gaussian functions.
This procedure ensures that the time offset would be 0 if the pulse profile of PSR~J1946+2052 were stable.
Finally, we used the Fourier domain algorithm of \cite{1992RSPTA.341..117T} with Markov chain Monte Carlo (FDM, implemented in the \texttt{pat} program of \texttt{PSRCHIVE}) to fit the phase offsets in each main pulse, and convert them to ToA offsets.

The result of the ToA offsets is displayed in Fig.~\ref{fig:offset}, wherein we also present the evolution of the half-width of the main pulse, denoted by width offsets.
The significant correlation between the ToA offsets and width offsets indicates how profile evolution will affect the ToA measurement.
Then we applied the linear relation between MJD and ToA offsets to all the ToAs we derived from FAST observations.

\begin{figure}
\begin{center}
\includegraphics[width=\columnwidth]{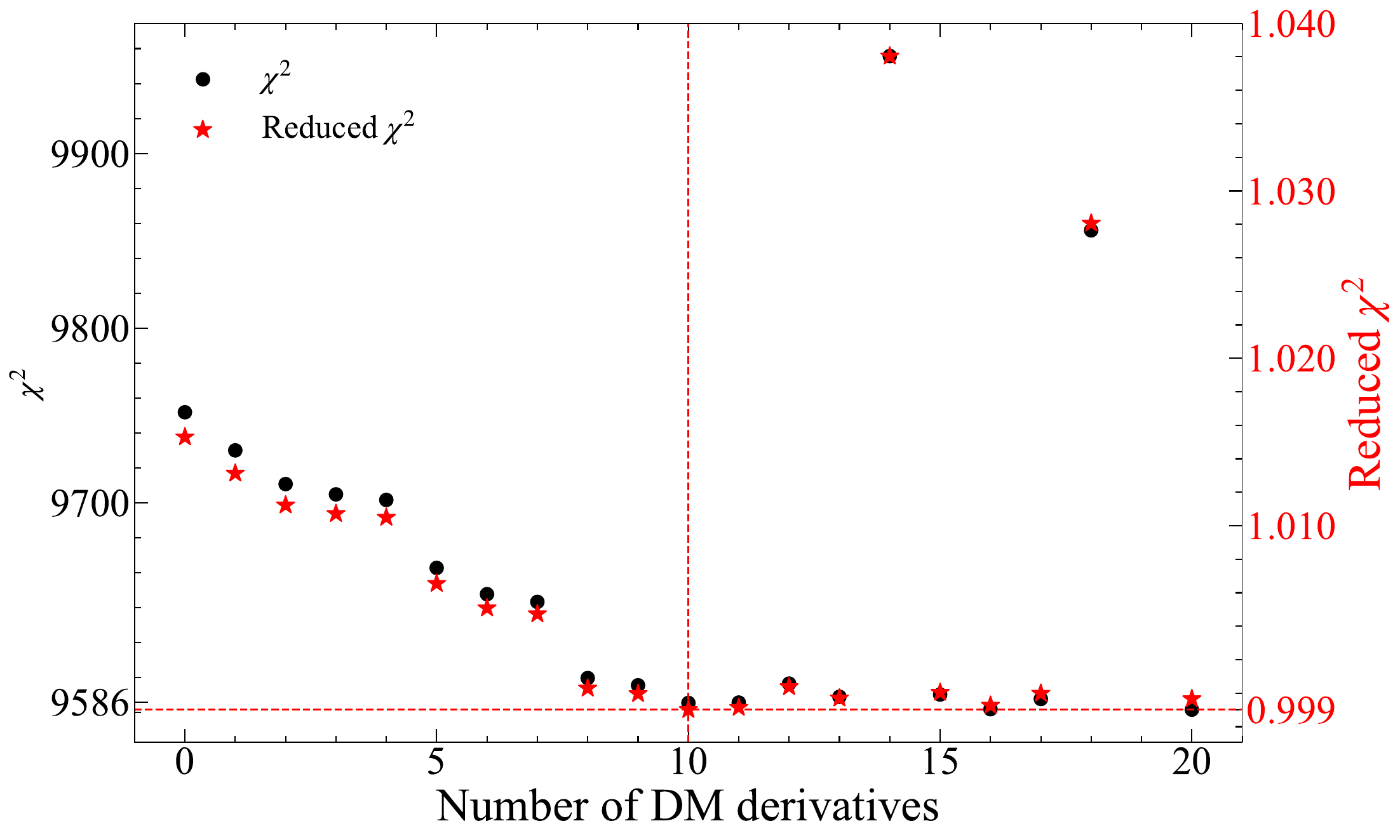}
\caption{The $\chi^2$ and reduced $\chi^2$ with different numbers of DM derivatives are shown in black solid circles and red solid stars, respectively. The 10-order DM derivative fit generates the lowest $\chi^2$ and reduced $\chi^2$. Using a higher order DM derivative will overfit, indicated by the larger $\chi^2$ and reduced $\chi^2$ after 10 DM derivatives.}
\label{fig:DMN}
\end{center}
\end{figure}

The pulse profile template from FAST observation uses a two-dimensional template that tracks the pulse profile as a function of both spin phase and radio frequency, whereas the Arecibo and GBT datasets rely on simpler one-dimensional templates.
Therefore, these template definitions do not share the exact same fiducial pulse longitude as a function of radio frequency; combining the data sets leaves small, systematic offsets in both pulse phase and DM.

Regarding the definition of the reference longitude on the NS, this problem is difficult to solve, given that the radio pulses at low frequency (like L-band) are significantly scatter-broadened relative to high frequency (S-band). If we use {\sc tempo} with more than 1 iteration, the phase offsets are translated to time offsets, which can bias the estimates of times of passage through periastron and thus of estimates of $P_{\rm b}$ and $\dot{P}_{\rm b}$ \citep[for a detailed discussion, see ][]{2021A&A...654A..16G}. This is not a problem in this work because we always ran {\sc tempo} with a single iteration. In this case, any offsets between ToA data sets are treated as phase offsets, originating from different definitions of the reference longitude of the NS. This can be done under the assumption that there are no significant offsets between the absolute timing of the different timing systems, which is known from other works on pulsars with better timing precision.

Regarding the DM offset between the FAST and Arecibo/GBT data, its exact value depends on the exact alignment of the pulse profiles at different frequencies in
the FAST 2-D template, which is affected by the DM that was used to create that profile. In Fig.~\ref{fig:DM_variation}, we display the DM measurements for a set of different epochs using the DMX model, a piecewise-constant function to describe the DM variation \citep{2013ApJ...762...94D}. The uncorrected measurements are shown in panel (a).
In this panel, the DM offset is very clear, and has a value of about 0.04$\, \rm pc\, cm^{-3}$.
After applying this DM offset onto the Arecibo and GBT ToAs (panel b), the DM variation during the entire observation period is still significant. This variation is mostly caused by the motion of the Earth and the pulsar: the radio waves travelling through the different paths from the pulsar to the Earth probe slightly different regions of the ionised interstellar medium (IISM), which will add to slightly different electron column densities. If the varying DM is not taken into account, it will introduce systematic uncertainties to the measurement of the pulsar properties. 
These DM variations can be used to probe the property of the IISM, which is shown in Appendix~\ref{sec:sf}.

We employed DM derivatives to describe these variations. The reason we chose DM derivatives rather than the DMX values in Fig.~\ref{fig:DM_variation} is to reduce the number of parameters we fit in the model.
To determine the number of DM derivatives to use, we compared the $\chi^2$ and the reduced $\chi^2$ from the fit of ToAs with different numbers of DM derivatives. The result is displayed in Fig.~\ref{fig:DMN}, which reveals that after the
10th DM derivative, the quality of the fit starts degrading, indicating over-fitting.
We also display the DM variation predicted by our best 10-DM derivative model in panel (b) of Fig.~\ref{fig:DM_variation}, which provides a good description for the measured DM variation.

Most of the results below are very robust, in the sense that they depend only weakly on the particular DM model used, with one exception, which is the measurement of the Shapiro delay parameters: given its small amplitude, the Shapiro delay is easily contaminated by systematic errors and thus depends significantly on the DM model. 
To deal with this issue, we decided in advance to report the solution with the number of derivatives that produces the lowest reduced $\chi^2$, without knowing in advance whether the values of the Shapiro delay parameters in that solution would match the predictions of GR or not.\footnote{
The F-test showed that adding a DM derivative will significantly improve the fitting, as long as the $\chi^2$ value reduces, which means the method could not provide a solid number of DM derivatives up to 20.}
All results reported below are based on that model.

We used two binary models to examine the results for timing analysis, both based on the binary model of Damour \& Deruelle \citep{damour1986general}.
The first is the ``DDFWHE'' model, which, like the standard Damour \& Deruelle (DD) model, is theory-independent. The only difference is that the Shapiro delay is parametrised differently; in this model the "orthometric parametrisation" is used instead~\citep{2010MNRAS.409..199F}; this is useful since it minimizes the correlation of Shapiro delay parameters, providing a better description of the possible location of the system in the mass-mass (or mass - inclination) planes, and providing, as we will see, improved tests of gravity theories. This model was implemented in {\sc tempo} by~\cite{2016ApJ...829...55W}.  These theory-independent models are necessary not only for understanding which relativistic effects are detected in the data but also to quantify them precisely and test gravity theories.

The second binary model, known as the ``DDGR'' model, is a theory-dependent model that assumes that GR provides a correct description of all relativistic effects in the system \citep{1989ApJ...345..434T}. In this model, the only two unknowns are the two masses, or in the specific formalism of the DDGR model, the total system mass ($M$\footnote{This is used in the model because, for highly eccentric systems, the periastron advance generally results in estimates of $M$ that are much more precise than either $m_{\rm p}$ or $m_{\rm c}$.}) and the companion mass ($m_{\rm c}$). 
Apart from immediate estimates of the masses, this model has the advantage of more easily detecting effects caused by, for instance, the acceleration of the system in the Galactic field, or alternatively the effects of spin-orbit coupling on the orbit of the system, because the difference between the observed value and the prediction from GR, such as in $\dot{P}_{\rm b}$ (XPBDOT) can be measured directly.

\section{Results}
\label{sec:results}

\begin{figure*}

\includegraphics[width=\textwidth]{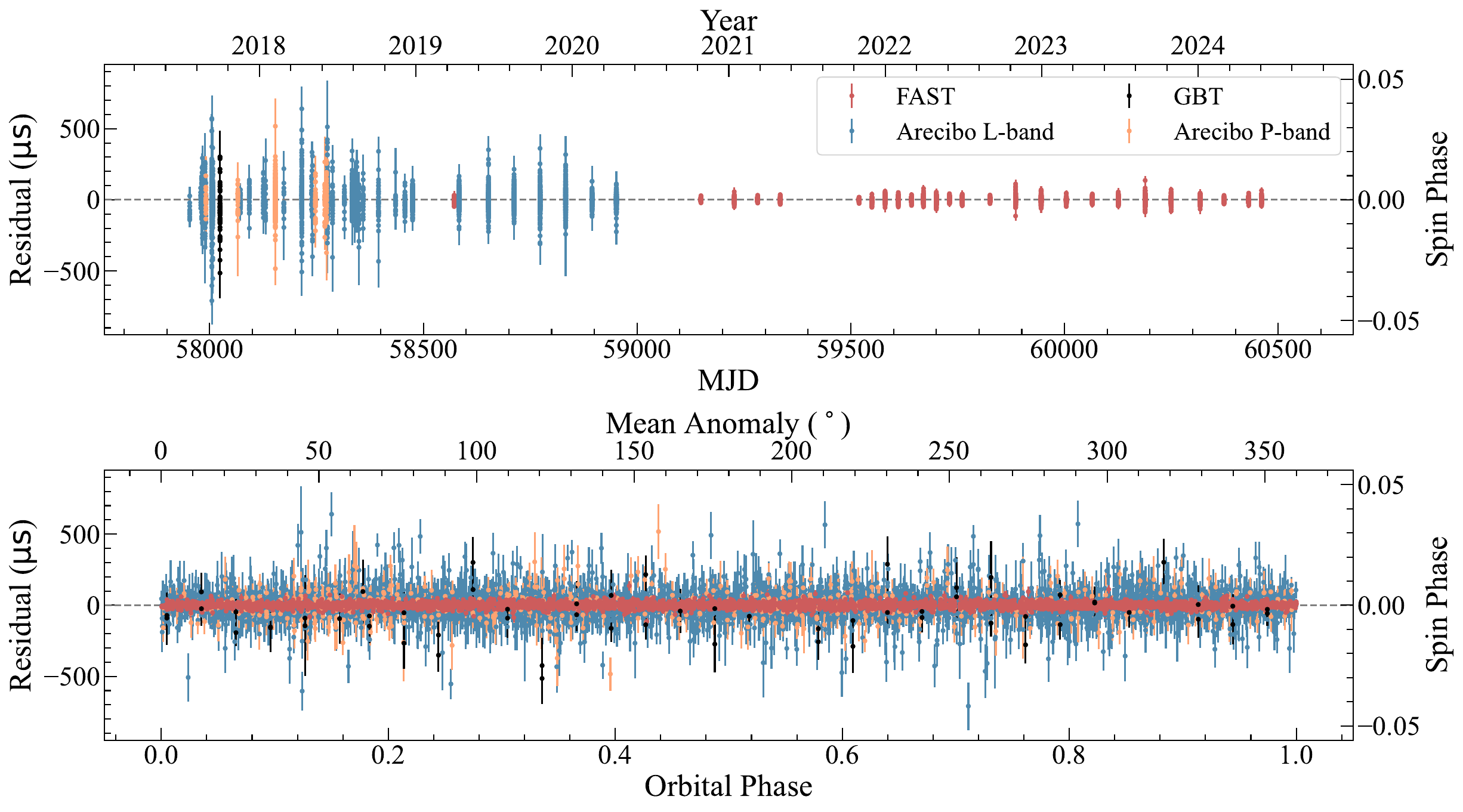}
\caption{Residuals obtained using the  DDFWHE timing solution in Table~\ref{tab:timsol}. The residuals in blue, orange, green and red are derived from L-band/PUPPI data, P-band/PUPPI data, single GBT observation and FAST data. Top: residuals as a function of time. Bottom: residuals as a function of orbital phase. 
The post-fit residuals' root-mean-square (RMS) is consistent with the ToA uncertainties, and the RMS of the residuals from Arecibo L-band, Arecibo P-band, GBT and FAST are 66.088$\,\upmu \rm s$, 87.709$\,\upmu \rm s$, 138.936$\,\upmu \rm s$ and 13.581$\,\upmu \rm s$, respectively. No unmodeled trends are seen in the ToA residuals, indicating that, within measurement uncertainty, the DDFWHE timing solution provides an adequate description of the timing of the system.}
\label{fig:residuals}

\end{figure*}

The parameters of our {\sc tempo} fits are presented in Table~\ref{tab:timsol}. The ToA residuals (ToA minus the prediction of the DDFWHE timing solution in that table) are
shown in Fig.~\ref{fig:residuals}. 
They display no observable trends, which implies that the DDFWHE solution provides, within their measurement accuracy, an adequate description of the observed ToAs.

\begin{table*}
  \renewcommand{\arraystretch}{0.8}
  \begin{center} 

  \caption{Fitted and derived parameters for PSR J1946$+$2052. Numbers in parentheses represent 1-$\sigma$ uncertainties from {\sc tempo}, scaled for reduced $\chi^2 = 1$. Numbers in square parentheses are derived from the DDGR model by assuming GR is correct.}
  \label{tab:timsol}
  \begin{tabular}{lcc}
  \hline\hline
  \multicolumn{3}{c}{Data and data reduction parameters} \\
  \hline
  Solar System ephemeris \dotfill & \multicolumn{2}{c}{DE440} \\
  Time Units \dotfill & \multicolumn{2}{c}{TDB} \\
  Clock \dotfill & \multicolumn{2}{c}{TT(BIPM2023)} \\
  
  Epoch (MJD) \dotfill & \multicolumn{2}{c}{57982.080242} \\
  Span of Timing Data (MJD) \dotfill & \multicolumn{2}{c}{57953--60460} \\
  Number of ToAs \dotfill & \multicolumn{2}{c}{9625} \\
  
  Binary model \dotfill & DDFWHE & DDGR \\
  RMS Residual ($\upmu \rm s$) \dotfill & 18.221 & 18.221 \\
  $\chi^2$ \dotfill & 9589.61 & 9589.64 \\
  Reduced $\chi^2$ \dotfill & 0.999& 0.997 \\
  \hline\hline
  \multicolumn{3}{c}{Measured parameters} \\
  \hline
  Right ascension, $\alpha$ (J2000.0) \dotfill & \multicolumn{2}{c}{19:46:14.13475(4)} \\
  Declination, $\delta$ (J2000.0) \dotfill & \multicolumn{2}{c}{+20:52:24.829(1)} \\
  Proper Motion in R.A., $\mu_\alpha$ (mas\,$\rm yr^{-1}$) \dotfill & \multicolumn{2}{c}{$-1.5(1)$} \\
  Proper Motion in Dec., $\mu_\delta$ (mas\,$\rm yr^{-1}$) \dotfill & \multicolumn{2}{c}{$-4.0(2)$} \\
  Pulse frequency, $\nu$ ($\mathrm{s^{-1}}$) \dotfill & \multicolumn{2}{c}{58.961654637293(4)} \\
  First derivative of pulse frequency, $\dot \nu$ ($\mathrm{s^{-2}}$) \dotfill & \multicolumn{2}{c}{$-$3.87147(3)$\times 10^{-15}$} \\
  Dispersion measure, DM ($\mathrm{pc\,cm^{-3}}$) \dotfill & \multicolumn{2}{c}{93.9281(9)} \\
  Number of DM derivatives \dotfill & \multicolumn{2}{c}{10}\\
  Orbital period, $P_\mathrm{b}$ (days) \dotfill & 0.07848805554(2) &  0.078488055530(8) \\
  Projected semi-major axis, $x$ (s) \dotfill & 1.154474(2) & 1.1544738(3) \\
  Orbital eccentricity, $e$ \dotfill & 0.0638363(8) & 0.0638365(4) \\
  Epoch of periastron, $T_\mathrm{0}$ (MJD) \dotfill & 57953.2123884(5) & 57953.2123886(2) \\
  Longitude of periastron, $\omega$ (deg) \dotfill & 130.357(2) & 130.3584(8) \\
  Rate of periastron advance, $\dot \omega$ ($\rm deg\,yr^{-1}$) \dotfill & 25.79205(40) &  [25.7918222] \\
  Observed change in orbital period, $\dot P_\mathrm{b, obs}$ ($10^{-12}\,\mathrm{s\;s^{-1}}$) \dotfill & $-$1.8296(15) & [$-$1.8286549] \\
  $\dot P_\mathrm{b, obs}$ minus GR prediction (XPBDOT), $\Delta\dot P_\mathrm{b}$ ($10^{-12}\,\mathrm{s\;s^{-1}}$) \dotfill & - & $-$0.0009(14) \\
  Gravitational Redshift and Time Dilation, $\gamma$ (ms) \dotfill & 0.2591(6) & [0.259029] \\
  Total System Mass, $M$ ($\mathrm{M_{\odot}}$) \dotfill & - & 2.531837(24) \\
  Companion Mass, $m_{\rm c}$ ($\mathrm{M_{\odot}}$) \dotfill & - & 1.2476(20) \\
  Orthometric ratio of Shapiro delay, $\varsigma$ \dotfill & 0.760(36) & - \\
  Orthometric amplitude of Shapiro delay, $h_3$ ($\upmu$s) \dotfill & 2.58(14) & - \\
  \hline\hline
  \multicolumn{3}{c}{Derived parameters} \\
  \hline
  Galactic longitude, $l$ (deg) \dotfill & \multicolumn{2}{c}{57.66} \\
  Galactic latitude, $b$ (deg) \dotfill & \multicolumn{2}{c}{$-$1.98} \\
  DM-derived distance (NE2001), $d_\mathrm{DM}$ (kpc) \dotfill & \multicolumn{2}{c}{4.2} \\
  DM-derived distance (YMW16), $d_\mathrm{DM}$ (kpc) \dotfill & \multicolumn{2}{c}{3.5} \\
  Galactic height (NE2001), $z_\mathrm{DM}$ (kpc) \dotfill & \multicolumn{2}{c}{0.16} \\
  Galactic height (YMW16), $z_\mathrm{DM}$ (kpc) \dotfill & \multicolumn{2}{c}{0.12} \\
  Magnitude of proper motion, $\mu$ ($\rm mas \, yr^{-1}$) \dotfill & \multicolumn{2}{c}{4.5(2)} \\
  Position angle of proper motion, $\Theta_{\mu}$ ($\deg$, J2000) \dotfill & \multicolumn{2}{c}{200(2)} \\
  Position angle of proper motion, $\Theta_{\mu}$ ($\deg$, Galactic) \dotfill & \multicolumn{2}{c}{260(2)} \\
  Heliocentric transverse velocity, $v_{\rm hel}$ ($\rm km \, s^{-1}$) \dotfill & \multicolumn{2}{c}{70(3)\tablefootmark{a}} \\
  Spin period, $P$ (s) \dotfill & \multicolumn{2}{c}{0.016960175323294(1)} \\
  Spin period derivative, $\dot P$ \dotfill & \multicolumn{2}{c}{1.113620(9)$\times 10^{-18}$} \\
  Inferred characteristic age, $\tau_\mathrm{c} = P/2 \dot{P}$ (Myr) \dotfill & \multicolumn{2}{c}{241} \\
  Inferred surface magnetic field, $B_\mathrm{S} = 3 \times 10^{19}\sqrt{ P \dot P}$ ($10^{9}$\,G) \dotfill & \multicolumn{2}{c}{4.4} \\
  Inferred light cylinder magnetic field, $B_{\rm LC}\simeq9.2 \, P^{-5/2}\dot{P}^{1/2} $ (G) \dotfill & \multicolumn{2}{c}{$8.2\times10^3$} \\
 Inferred spin-down luminosity, $\dot{E}\simeq0.395\,\dot{P}P^{-3}$ ($10^{33}\,\rm erg\,s^{-1}$)\dotfill  & \multicolumn{2}{c}{9\tablefootmark{b}} \\
  Mass function, $f_\mathrm{mass}$ ($\mathrm{M_{\odot}}$) \dotfill & 0.268181(1) & 0.2681809(2) \\
  Total System Mass, $M$ ($\mathrm{M_{\odot}}$) \dotfill & 2.531858(60)\tablefootmark{c} & 
  - \\
  Companion Mass, $m_{\rm c}$ ($\mathrm{M_{\odot}}$) \dotfill & 1.2480(21)\tablefootmark{d} & - \\
  Pulsar mass, $m_{\rm p}$ ($\mathrm{M_{\odot}}$) \dotfill & 1.2838(21) & 1.2842(21) \\
  Sine of orbital inclination, $\sin(i)$ \dotfill & 0.9599(16)\tablefootmark{e} & - \\
  \hline

\end{tabular} 
\tablefoot{Numbers in square parentheses are derived from the DDGR model by assuming GR is correct.\\
\tablefoottext{a}{Used the DM derived distance from the YMW16 model.}\\
\tablefoottext{b}{The expression used to calculate this assumes the moment of inertia is $I = 1\times 10^{45}\, \rm g\, cm^2$.}\\
\tablefoottext{c}{Derived from $\dot{\omega}^{\rm 1PN}$.}\\
\tablefoottext{d}{Derived from $\dot{\omega}^{\rm 1PN}$ and $\gamma$.}\\
\tablefoottext{e}{Derived from $\dot{\omega}^{\rm 1PN}$, $\gamma$ and $f_\mathrm{mass}$.}\\
}
\end{center}

\end{table*}

We will now discuss some of the parameters in this timing solution that will be especially relevant for the following discussions.

\subsection{Position and Proper motion}

We have improved the precision of the position measurement by combining TOAs from FAST, Arecibo and GBT.
The timing solution provides a position with R.A. (J2000) 19:46:14.13475(4) and DEC. (J2000) +20:52:24.829(1) on 2017 Aug. 17.
Combining the position with the DM, we are able to estimate the distance to the pulsar as $d = 4.2$\,kpc based on the NE2001 model \citep{2002astro.ph..7156C} and 3.5\,kpc based on the YMW16 model \citep{2017ApJ...835...29Y}.
The result indicates a relatively low galactic height of less than 0.2 kpc, suggesting that the system is not likely to have a large vertical velocity.

After 7 years of timing, we are able to constrain the proper motion with high precision: $-$1.5(1)\,mas\,$\rm yr^{-1}$ in R.A. and $-$4.0(2)\,mas\,$\rm yr^{-1}$ in Dec., leading to a total proper motion of $\mu = 4.2(2)\rm \,mas\,yr^{-1}$.
With this proper motion and the distance from the YMW16 model, we derive the heliocentric transverse velocity of the system of $v_{\rm T} = 70(3)\rm \,km\,s^{-1}$. The position angle of the proper motion, in Galactic coordinates, is $\Theta_\mu \, = \, 260 \pm 2\, \deg$, which is 10 degrees South from the Western direction of the Galactic plane. 
The implications of this measurement for the evolution of the system will be discussed in a future publication.
For what follows, the value of $\mu$ will be the most used.

\subsection{Spin period derivative}
\label{sec:spin derivative}

\begin{table}[]
    \centering
        \caption{The observed $\dot{P}$, the contribution of the Galactic acceleration and the Shklovskii effect and the intrinsic $\dot{P}$.}
    \begin{tabular}{c c c}
    \hline
    \hline
            &  YMW16  &  NE2001 \\[0.1ex]
            &  \multicolumn{2}{c}{(s\,$\rm s^{-1}$)}  \\ 
    \hline
    $\dot{P}_{\rm obs}$  &  \multicolumn{2}{c}{$1.113620(9)\times10^{-18}$}\\
    
    $\dot{P}_{\rm gal}$  &  $-4.8(16)\times10^{-21}$
                         &  $-5.9(20)\times10^{-21}$\\
    $\dot{P}_{\rm shk}$  &  $+2.63(58)\times10^{-21}$
                         &  $+3.15(70)\times10^{-21}$\\
    $\dot{P}_{\rm int}$  &  $+1.1158(17)\times10^{-18}$
                         &  $+1.1164(21)\times10^{-18}$\\
    \hline
    \end{tabular}

    \label{tab:pdot_table}
\end{table}

The measurement of the spin period derivative has been improved by four orders of magnitude.
Its value is 1-$\sigma$ consistent with the measurement in Paper~2018,
$0.9(2) \times 10^{-18}$. This and the measurement of $\mu$ provide us with a chance to estimate the intrinsic spin-down rate.
The observed spin-down rate is modified from the intrinsic one by the unknown radial velocity of the pulsar system in the form of the Doppler factor ($D$), as given by \citep{1993ASPC...50..141P}:
\begin{equation}
    P_{\rm{obs}} = D^{-1}P_{\rm{int}} = \left[1+(\mathbf{ V_{\rm{PSR}}} - \mathbf{{V}_{\rm{SSB}}})\cdot\mathbf{n}/c \right]^{-1}P_{\rm int},
\end{equation}
where $\mathbf{V_{\rm PSR}}$ is the pulsar system velocity, $\mathbf{V_{\rm SSB}}$ is the velocity of the Solar System Barycentre (SSB), $\mathbf{n}$ is the unit vector pointing from the SSB to the pulsar system, $P_{\rm obs}$ is the observed spin period and $P_{\rm int}$ is the intrinsic spin period.
By differentiating the equation above, one gets \citep{1993ASPC...50..141P}:
\begin{equation}
    \dot{P}_{\rm obs} = \dot{P}_{\rm int}
                        +\dot{P}_{\rm gal}
                        +\dot{P}_{\rm shk}
    \label{eq:accel}
\end{equation}
where the second term results from the differential Galactic acceleration between the Solar and the pulsar system, and the third term is the Shklovskii term \citep{1970SvA....13..562S}. The last two terms can be calculated according to
\begin{equation}
    \frac{\dot{P}_{\rm gal}}{P} =
    -\frac{(\mathbf{a_{\rm PSR}-\mathbf{a_{\rm SSB}})}\cdot\mathbf{n}}{c}
    \quad\mbox{and}\quad
    \frac{\dot{P}_{\rm shk}}{P}=-\frac{\mu^2d}{c},
    \label{eq:accel_2}
\end{equation}
where $\mathbf{a_{\rm PSR}}$ and $\mathbf{a_{\rm SSB}}$ are the accelerations of the pulsar system and the SSB in the gravitational potential of the Milky Way (see also \citealt{1991ApJ...366..501D}).

To calculate the accelerations of the pulsar system and the SSB, we adopted the Galactic potential model \texttt{MWPotential2014} deployed in the Python package \texttt{galpy}\footnote{http://github.com/jobovy/galpy} \citep{2015ApJS..216...29B}, scaled such that the distance from the Sun to the Galactic centre is $R_0 = 8.275(34)\,\rm kpc$ \citep{2021A&A...647A..59G}, and the circular velocity of the Sun's local standard of rest is $\Theta_0 = 240.5(41)\,\rm km\,s^{-1}$ \citep{2021A&A...654A..16G}.
For our calculations of the differential Galactic acceleration, we further assumed the Sun's height above the local Galactic mid-plane as $Z_\odot=20.8(3)\,\rm pc$ \citep{2019MNRAS.482.1417B}.

Then we used the two distances determined by YMW16 and NE2001 as Gaussian distributions with 20\% uncertainty \citep{2002astro.ph..7156C} and derived the contribution from the Galactic acceleration and the Shklovskii effect. The results are presented in Table~\ref{tab:pdot_table}. 
The resulting $\dot{P}_{\rm int}$ values from the two DM models are consistent with each other as $1.116(2)\times10^{-18}\, \rm s\,s^{-1}$, and are larger than $\dot{P}_{\rm obs}$ by a factor of 1.002(1).
Using the expressions in \cite{lorimer2005handbook}, this results in a characteristic age of 0.24 Gyr, a B-field of $4.4\, \times \, 10^9 \, \rm G$ and, by using a moment of inertia as $I = 1.31\times 10^{45} \, \rm g\, cm^2$ (see in Appendix~\ref{sec:appendixA}), a spin-down energy of $1.183(2) \times 10^{34}\, \rm erg \, s^{-1}$.

\subsection{Rate of advance of periastron}
\label{advance}

As mentioned earlier, the defining feature of PSR J1946+2052 is its extremely short orbital period,  $P_{\rm b} = 0.0785 \, \rm{d}$. This parameter primarily accounts for the fact that, as outlined in Paper~2018, this binary pulsar exhibits the highest recorded rate of periastron advance, $\dot{\omega}\, = \, 25.6(3) \, \rm \deg \, yr^{-1}$.

This parameter has been massively improved in this work: our current measurement,
$\dot{\omega}\, = \, 25.79205(40) \, \rm \deg \, yr^{-1}$. This is 1-$\sigma$ compatible with the measurement presented in Paper~2018,
but three orders of magnitude more precise. Assuming that this is as predicted by GR, then the total mass of the system in solar mass parameters is, to leading post-Newtonian order, given by \cite{Robertson_1938}:
\begin{equation}
M = \frac{1}{T_{\odot}} \left[ \frac{\dot{\omega}}{3} (1 -e^2) \right]^{\frac{3}{2}} n_\mathrm{b}^{-\frac{5}{2}}, \ \ \ n_\mathrm{b} \equiv \frac{2 \pi}{P_\mathrm{b}},
\label{eq:mtot}
\end{equation}
where $T_{\odot} \equiv ({\cal G M})_{\odot}^{\rm N} / c^3
= 4.925490947641266978...\,\upmu \rm s$ is an exact quantity, the nominal solar mass parameter
$({\cal G M})_{\odot}^{\rm N}$ in time units \citep{2016AJ....152...41P}\footnote{In the equations containing $T_{\odot}$, the mass values are adimensional, expressing the ratio $Gm / ({\cal G M})_{\odot}^{\rm N}$, where $m$ is the corresponding mass in mass units. Explicit mass values in the text are followed by the symbol M$_{\odot}$ to indicate that they are multiples of the solar mass parameter.} and $e$ is the orbital eccentricity.
Using the values for these quantities from Table~\ref{tab:timsol}, we obtain
$M \, = \, 2.531858(60) \, \rm M_{\odot}$. 
This is the lowest total mass measured among known DNS systems in the Galactic field, with the possible exception of PSR~J1411+2551 ($M \, = \, 2.538(22) \, \rm M_{\odot}$, \citealt{2017ApJ...851L..29M}).
Also, as noted in the Introduction, this DNS has the lowest eccentricity for any known
DNS in the Galaxy.

\subsection{Einstein delay}
Whereas the timing solution in Paper~2018 is sensitive to only one PK parameter ($\dot{\omega}$), our enhanced timing precision and extended timeline of the data set enabled the detection of four additional PK parameters.
The first is the Einstein delay amplitude, $\gamma = 0.2591 \pm 0.0006\, \rm ms$, which is $\sim 430$-$\sigma$ significant. In GR, this is given by:
\begin{equation}
\gamma = g \, \frac{m_{\rm c} (M + m_{\rm c}) }{M^{\frac{4}{3}}},
\label{eq:gamma}
\end{equation}
where
\begin{equation}
g = e \, n_\mathrm{b}^{-\frac{1}{3}} T_{\odot}^{\frac{2}{3}},
\end{equation}
and where $m_{\rm c}$ is the companion mass.
Although the small values of $e$ and $n^{-\frac{1}{3}}_\mathrm{b}$ make our value of $g$ the smallest for any DNS system, which could result in a large relative uncertainty, our measurement of $\gamma$ is still extremely precise with an uncertainty of 600 ns, owing to the large precession angle ($\Delta  \omega = T \dot{\omega} = 184.09\, \deg$) that has been covered over the length of our timing baseline ($T$). 

With a measurement of $M$ and $\gamma$, we can make a first estimate of the individual masses of the
components. The mass of the companion is given by  \citep{2019MNRAS.490.3860R}:
\begin{equation}
m_{\rm c} = \frac{1}{2} \left( \sqrt{M^2 + 4M^{\frac{4}{3}}\frac{\gamma}{g}} - M\right),
\label{eq:gamma_masses}
\end{equation}
from which we get $m_{\rm c} = 1.2480(21) \, \rm M_{\odot}$. The mass of the pulsar is given by $m_{\rm p} = M - m_{\rm c} = 1.2838(21) \, \rm M_{\odot}$. This indicates that PSR~J1946+2052 is a slightly asymmetric system.
Rewriting the mass function equation, we obtain
\begin{equation}
\sin i = \frac{x}{m_{\rm c}} \, T_{\odot}^{-\frac{1}{3}} \left( n_{\rm b} M \right)^{\frac{2}{3}},
\label{eq:sini}
\end{equation}
where $i$ is the orbital inclination.
From this we obtain $\sin i = 0.9599(16)$, which corresponds to
$i = 73.71(33) \, \deg$ or $i = 106.29(33) \, \deg$. This is 2-$\sigma$ consistent with
$i = {63^\circ}^{+5^\circ}_{-3^\circ}$ derived by \cite{2024ApJ...966...46M}, which includes the small
misalignment between the spin axis of the pulsar and the orbital angular momentum that they estimated.

\subsection{Shapiro delay}
\label{subsec:shapiro}

In our timing, we also detect, for the first time, the Shapiro delay. To quantify this detection, we adopted the orthometric parametrisation introduced by \cite{2010MNRAS.409..199F}.
The orthometric amplitude $h_3$ and the orthometric ratio $\varsigma$ of Shapiro delay are given in GR by
\begin{equation}
    \varsigma=\frac{\sin i}{1+\sqrt{1 - \sin^2i}},\ \ \ h_3= m_{\rm c} T_{\odot}\,\varsigma^3.
    \label{eq:shapiro}
\end{equation}
The advantage of using such a parametrisation relative to the $r$, $s$ parametrisation in the DD model is that it reduces the correlation between the two parameters of the Shapiro delay. Furthermore, as we will see below, the $h_3$ test also provides a more precise test of GR compared to the $r$ parameter used in the DD model.

This parametrisation has been implemented in the DDFWHE timing model in \texttt{TEMPO}. The values measured from the timing of PSR~J1946+2052 derived the values of $h_3 =2.58(14)\,\upmu \rm s$ and $\varsigma = 0.760(36)$, i.e., this is a highly significant detection of the Shapiro delay. From these, we derive $\sin i = 0.963(13)$ and $m_{\rm c} = 1.21(19)\,\rm M_{\odot}$. These values are consistent with the values derived above from $\dot{\omega}$ and $\gamma$, but less precise.
The larger deviation from $i = 90^\circ$ implies that the Shapiro delay signal is much weaker than in the Double Pulsar
\citep{2021PhRvX..11d1050K}, hence the much larger relative uncertainties in the measurement of the Shapiro delay parameters.

Using the DD model, we obtain $s = 0.962(13)$ and $m_{\rm c, S} = r/T_{\odot} = 1.21(21)$, which
are again consistent with the values derived from the DDFWHE solution.
The values of $i$ derived from the $\varsigma$ in the DDFWHE solution and $s$ in the DD solution are both 
$i = 74 \pm 2 \, \deg$. This is not surprising given the exclusive dependence of $\varsigma$ and $s$ on $i$, but is important for the interpretation of the GR tests made with $\varsigma$ and $s$.

\subsection{Orbital decay}
\label{sec:pbdot}

\begin{figure}
    \centering
    \includegraphics[width=1.\linewidth]{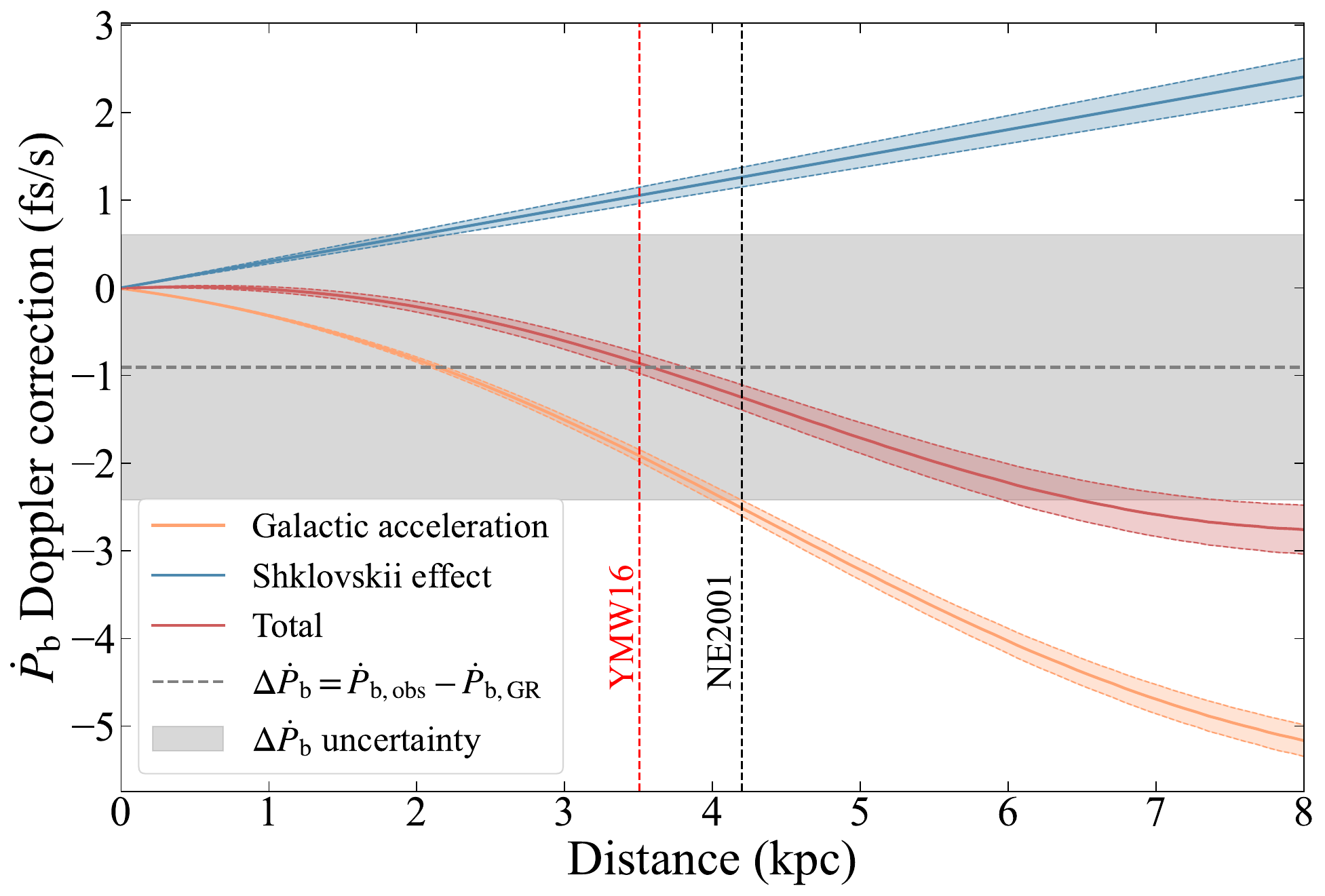}
    \caption{$\dot{P}_{\rm b}$ contributions from differential Galactic acceleration and the Shklovskii effect as a function of the distance to the pulsar (see text for details). The orange and blue areas indicate the 1-$\sigma$ confidence of these two effects, the solid lines represent the nominal values and the dotted curves the $\pm 1$-$\sigma$ uncertainties. The total external contribution of $\dot{P}_{\rm b}$ is shown by the corresponding red area and curves. The uncertainties come from the uncertainty of $R_0$, $\Theta_0$ and the proper motion. The dashed red and black lines are the distances derived from YMW16 and NE2001. 
    The grey area displays $\Delta \dot{P}_{\rm b} = \dot{P}_{\rm b, obs} - \dot{P}_{\rm b, GR}$. This quantity, estimated in detail in Sect.~\ref{sec:masses_bayesian}, represents a measurement of the total external contribution that assumes the validity of GR; its width represents its uncertainty, which is dominated by the error of the observed $\dot{P}_{\rm b}$.}
    \label{fig:enter-label}
\end{figure}

\begin{table}[]
    \centering
    \caption{Terms associated with the variation of the orbital period.
    }
    \begin{tabular}{l c c}
    \hline
    \hline
            &  YMW16  &  NE2001 \\[0.1ex]
            &  \multicolumn{2}{c}{(fs\,$\rm s^{-1}$)}  \\ 
    \hline
    $\dot{P}_{\rm b, obs}$  &  \multicolumn{2}{c}{$-1829.6(15)$}\\
    $\dot{P}_\mathrm{b, GR}$  &  \multicolumn{2}{c}{$-1828.7(1)$} \\
    $\Delta \dot{P}_{\rm b}$  &  \multicolumn{2}{c}{$-0.9^{+14}_{-15}$} \\    
    $\dot{P}_{\rm b, gal}$  &  $-1.93(62)$
                                 &  $-2.46(68)$\\
    $\dot{P}_{\rm b, shk}$  &  $+1.05(23)$
                                 &  $+1.25(28)$\\
     $\dot{P}_{\rm b, ext}$  &  $-0.88(66)$
                                 &  $-1.21(74)$\\
    $\dot{P}_{\rm b, int}$  &  $-1828.8(16)$
                                 &  $-1828.4(16)$ \\
    $\dot{P}_\mathrm{b, xs}$  & $-0.1(16)$ & $+0.3(16)$ \\
    \hline
    \end{tabular}
\tablefoot{The term $\Delta \dot{P}_{\rm b} = \dot{P}_{\rm b, obs} - \dot{P}_\mathrm{b, GR}$ represents a constraint on the sum external contributions to $\dot{P}_{\rm b}$, this is calculated in detail in Sect.~\ref{sec:masses_bayesian}; this value is represented by the gray bar in Fig.~\ref{fig:enter-label}. The sum of the external contributions calculated by our models for the two assumed distances, $\dot{P}_{\rm b, ext} = \dot{P}_{\rm b, gal} + \dot{P}_{\rm b, shk}$, are in good agreement with $\Delta \dot{P}_\mathrm{b}$ (see red band in Fig.~\ref{fig:enter-label}). The intrinsic $\dot{P}_{\rm b}$, $\dot{P}_{\rm b, int}$ is defined as $\dot{P}_{\rm b, obs} - \dot{P}_{\rm b, gal} - \dot{P}_{\rm b, shk}$; if GR is correct, this should match  $\dot{P}_\mathrm{b, GR}$. The difference between $\dot{P}_{\rm b, int}$ and $\dot{P}_\mathrm{b, GR}$ is the ``excess'' variation of the orbital period, $\dot{P}_{\rm b, xs}$ which quantifies possible deviations from GR.}
    
    \label{tab:pbdot_table}
\end{table}

As a result of our timing, we now have a 1200-$\sigma$ significant measurement or the observed variation of the orbital period: $\dot{P}_{\rm b, obs} = -1829.6 \, \pm \, 1.5\,\rm fs/s$. 

Most of this is due to orbital decay induced by GW damping. In GR, this orbital decay ($\dot{P}_{\rm b, GR}$) is given by \citep{Peters_1964}:
\begin{equation}
    \label{eq:pbdot}
    \dot{P}_{\rm b, GR}=-\frac{192\pi}{5} \left( T_{\odot} n_b \right)^{\frac{5}{3}} f(e)\frac{m_{\rm c} (M - m_{\rm c}) }{M^{\frac{1}{3}}},
\end{equation}
where 
\begin{equation}
    f(e)=\frac{1+(73/24)e^2+(37/96)e^4}{(1-e^2)^{\frac{7}{2}}};
\end{equation}
these expressions were re-written as a function of $M$ because of its high precision, but also because $m_{\rm c}$ and $m_{\rm p}$ are not determined independently.

Like $\dot{P}_{\rm obs}$, the value of $\dot{P}_{\rm b, obs}$ is affected by the variation of the Doppler factor exactly the same way as $\dot{P}$, i.e., with contributions from the relative Galactic accelerations ($\dot{P}_{\rm b, gal}$) and the Shklovskii effect ($\dot{P}_{\rm b, shk}$).
Therefore, we can modify Eq.~(\ref{eq:accel}) as:
\begin{equation}
    \dot{P}_{\rm b, obs}=\dot{P}_{\rm b, int} + \dot{P}_{\rm b, ext}=\dot{P}_{\rm b, int}+\dot{P}_{\rm b, gal} + \dot{P}_{\rm b, 
    shk},
    \label{eq:pbdot_contributions}
\end{equation}
where $\dot{P}_{\rm b, int}$ is the intrinsic variation of the orbital period, which is dominated by the emission of GW, $\dot{P}_{\rm b, ext}$ is the external contribution of the orbital decay, and
\begin{equation}
    \dot{P}_{\rm b, gal} = \frac{\dot{P}_{\rm gal}}{P} P_{\rm b},\ \  \dot{P}_{\rm b, shk}=\frac{\dot{P}_{\rm shk}}{P} P_{\rm b}.
    \label{eq:pbdot_pdot}
\end{equation}
As in Sect.~\ref{sec:spin derivative}, we used the distances from YMW16 and NE2001 and assumed a 20\% uncertainty. Considering the uncertainty of the proper motion and using the distance from YMW16, $\dot{P}_{\rm b,gal}$ and $\dot{P}_{\rm b, shk}$ can be determined as $-1.93(62)\,\rm fs\,s^{-1}$ \footnote{It is important to note that the calculated $\dot{P}_\mathrm{b,gal}$ is (slightly) different when using different Galactic potential models, and that all commonly used models are only a rough approximation to the true gravitational potential of our Galaxy (see e.g.\ \cite{2019MNRAS.482.3249Z}, \cite{2021A&A...654A..16G}, \cite{Donlon_2024}). Fortunately, PSR J1946+2052 is at a very low Galactic latitude, in a region where such models are expected to give a good approximation to the (rather flat) rotation curve of the Galaxy. Furthermore, the uncertainty in $\dot{P}_\mathrm{b,gal}$ is clearly dominated by the uncertainty in the distance, and is still considerably smaller than the uncertainty in the observed $\dot{P}_\mathrm{b}$. Nevertheless, in addition to {\tt MWPotential2014} (Bovy 2015), which we have used in the main text, we have also performed calculations based on {\tt McMillan17} \citep{McMillan_2017} and {\tt Cautun20} \citep{2020MNRAS.494.4291C}, which give $-1.88(53)~\mathrm{fs\,s^{-1}}$ and $-1.84(53)~\mathrm{fs\,s^{-1}}$ respectively. These values are clearly consistent with the other values used here.}.
and $1.05(23)\rm \, fs\,s^{-1}$ respectively.

The sum of the external effects (for this distance and a 20\% uncertainty) is thus given by $-0.88(66)\,\rm fs\,s^{-1}$, which is consistent with $\Delta \dot{P}_{\rm b} = \dot{P}_{\rm b} - \dot{P}_{\rm b, GR}$ calculated in Sect.~\ref{sec:masses_bayesian}, the latter is indicated by the gray bar in Fig.~\ref{fig:enter-label}. The precision of this sum is mostly limited by the uncertainty in the distance, although the uncertainty of the proper motion remains important.

This implies that $\dot{P}_{\rm b, int} = -1828.8(16)\,\rm fs\,s^{-1}$, which is consistent with $\dot{P}_{\rm b, GR}$. The precision of this parameter is mostly limited by the precision of $\dot{P}_{\rm b, obs}$, but this will improve fast in the near future. All $\dot{P}_{\rm b}$ values derived assuming the NE2001 distance are 1-$\sigma$ consistent; they are presented in Table~\ref{tab:pbdot_table}.

Another contribution that could change the orbital period is the mass loss due to the spin-down of the pulsar. \cite{1991ApJ...366..501D} estimate it as:
\begin{equation}
\label{eq:pbdot_massloss}
    \frac{\dot{P}_{\rm b}^{\dot{m}_{\rm p}}}{P_{\rm b}}=
    8\pi^{2}\frac{I_{\rm p}\dot{P}_{\rm int}}{Mc^2P^3},
\end{equation}
where $I_{\rm p}$ is the moment of inertia of the pulsar.
By using the total mass derived in Sect.~\ref{advance} and the timing parameters in Table~\ref{tab:timsol}, $\dot{P}_{\rm b}^{\dot{m}_{\rm p}}$ can be calculated as $3.5(2)\times10^{-17}\, \rm s\,s^{-1}$ by using $I_\mathrm{p} = (1.31 \pm 0.08) \times 10^{45}~\mathrm{g\,cm^2}$ from Appendix~\ref{sec:appendixA}, which is completely ignorable in the current analysis in the orbital decay.

Owing to the short spin period (17\,ms) and the small magnetic field strength at the surface ($4.4\times10^{9}\,\rm G$), it is clear that the pulsar is the first-born NS in this binary system and the companion is the second-born NS.
Calculating the orbit into the past, we can see there were a few times when it crossed the Galactic plane within the 
characteristic age of the pulsar. By assuming the system formed at the time of the last crossing, we can estimate a conservative lower limit of the characteristic age of the companion with ${\rm tan}\,b/{\mu_b} \approx10\,\rm Myr$, where $b$ is the Galactic latitude and $\mu_b$ is the Galactic latitude component of the proper motion.

Since the companion is the second-born (hence non-recycled) NS, the location of the companion in the $P-\dot{P}$ diagram will be in the region of normal pulsars.
Therefore, we can simply estimate the lower limit of the companion's spin period ($P_{\rm c}$) and the first derivative of the spin period ($\dot{P}_{\rm c}$) by using the characteristic age calculated above in the $P-\dot{P}$ diagram \citep{lorimer2005handbook}, resulting in $P_{\rm c}\ge 180\,\rm ms$ and $\dot{P}_{\rm c} \approx 6\times10^{-15}\,\rm s\,s^{-1}$.
Utilising Eq.~(\ref{eq:pbdot_massloss}), the upper limit on the change of the orbital period caused by the mass loss of the companion could be estimated as $1.8\times10^{-20}\,\rm s\,s^{-1}$, which can be completely neglected in the current analysis of the orbital decay.

\section{Masses and tests of general relativity}
\label{sec:masses}

\begin{figure*}
\sidecaption
    \includegraphics[width=12cm]{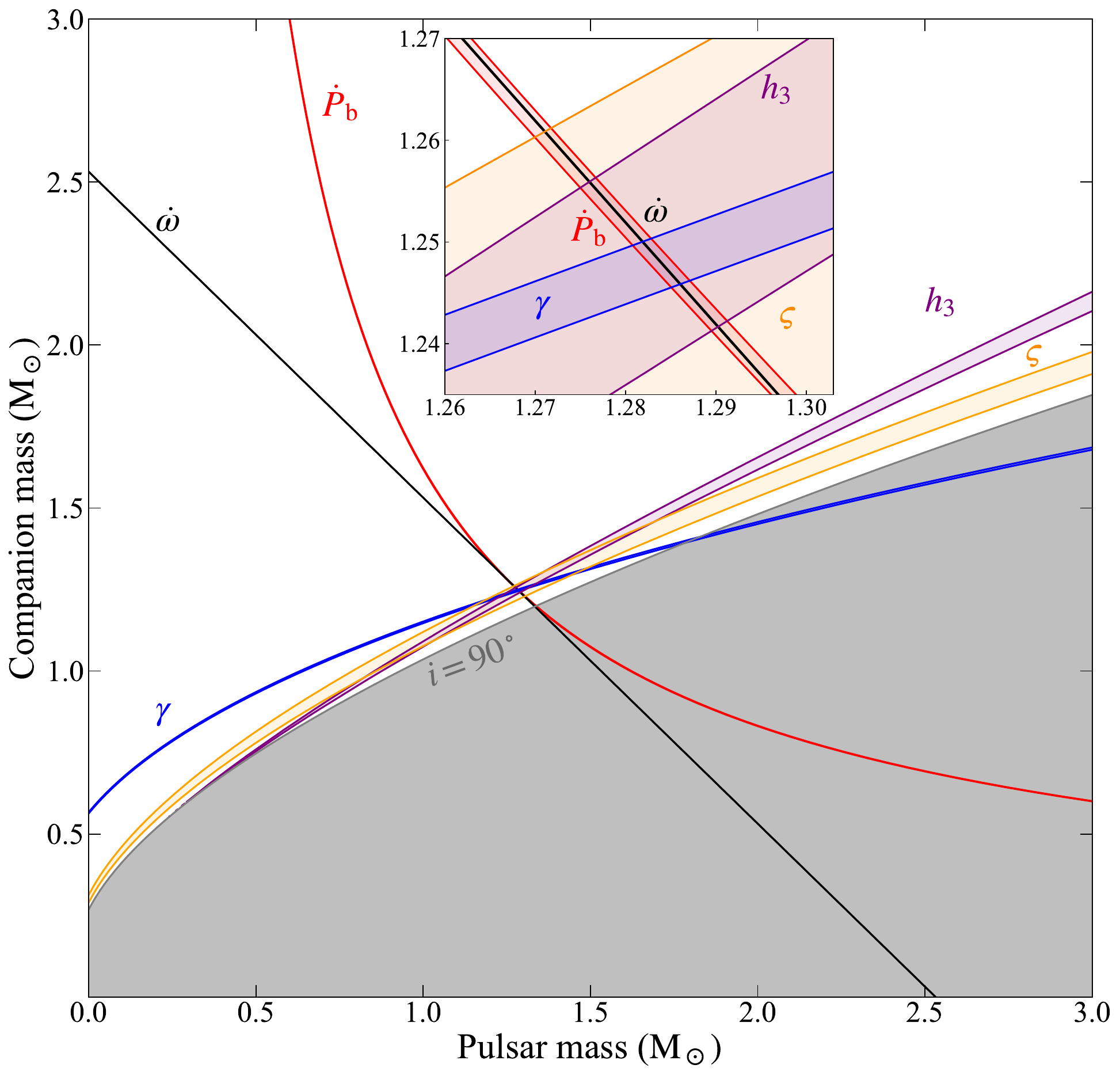}
    \caption{Mass-mass diagram for PSR~J1946+2052. The regions consistent with the measured $\dot{P}_{\rm b},\ \dot{\omega},\ \gamma,\ h_3,\ \varsigma$ and their 1-$\sigma$ uncertainties are displayed in red, green, blue, purple and orange. Note that $\dot{P}_{\rm b}$ in this diagram is corrected by removing the external contributions. The grey area is excluded by the mass function and $\sin i \leq 1$.
    The inset is an expanded view, showing in more detail the intersection of all PK parameters and the self-consistency of GR. 
    The $\dot\omega$-line is based on the leading-order equation~(\ref{eq:mtot}). The (small) modifications related to the next-to-leading order contributions to the periastron advance are discussed in detail in Sec.~\ref{sec:omdot_NLO}.
    }
    \label{fig:mmdiagram}
\end{figure*}

The measurement of 5 PK parameters enables us to determine the masses of the pulsar and the companion, and 
perform three tests of GR by testing the self-consistency of Eqs.~(\ref{eq:mtot}), (\ref{eq:gamma}), (\ref{eq:shapiro}) and (\ref{eq:pbdot}) for the Keplerian and PK parameters of the system.

This consistency is depicted graphically in the mass-mass diagram in Fig.~\ref{fig:mmdiagram}. In this diagram, we highlight the regions where the masses are 1-$\sigma$ consistent with the measurements of $\dot{\omega}$,\ $\gamma$,\ $h_3$,\ $\varsigma$ and $\dot{P}_{\rm b}$ using the aforementioned equations; these form bands in the diagram. As we can see, all bands meet at a small common region perfectly even in the enlarged panel, indicating GR passed the three tests.

\begin{figure*}
\sidecaption
    \includegraphics[width=12cm]{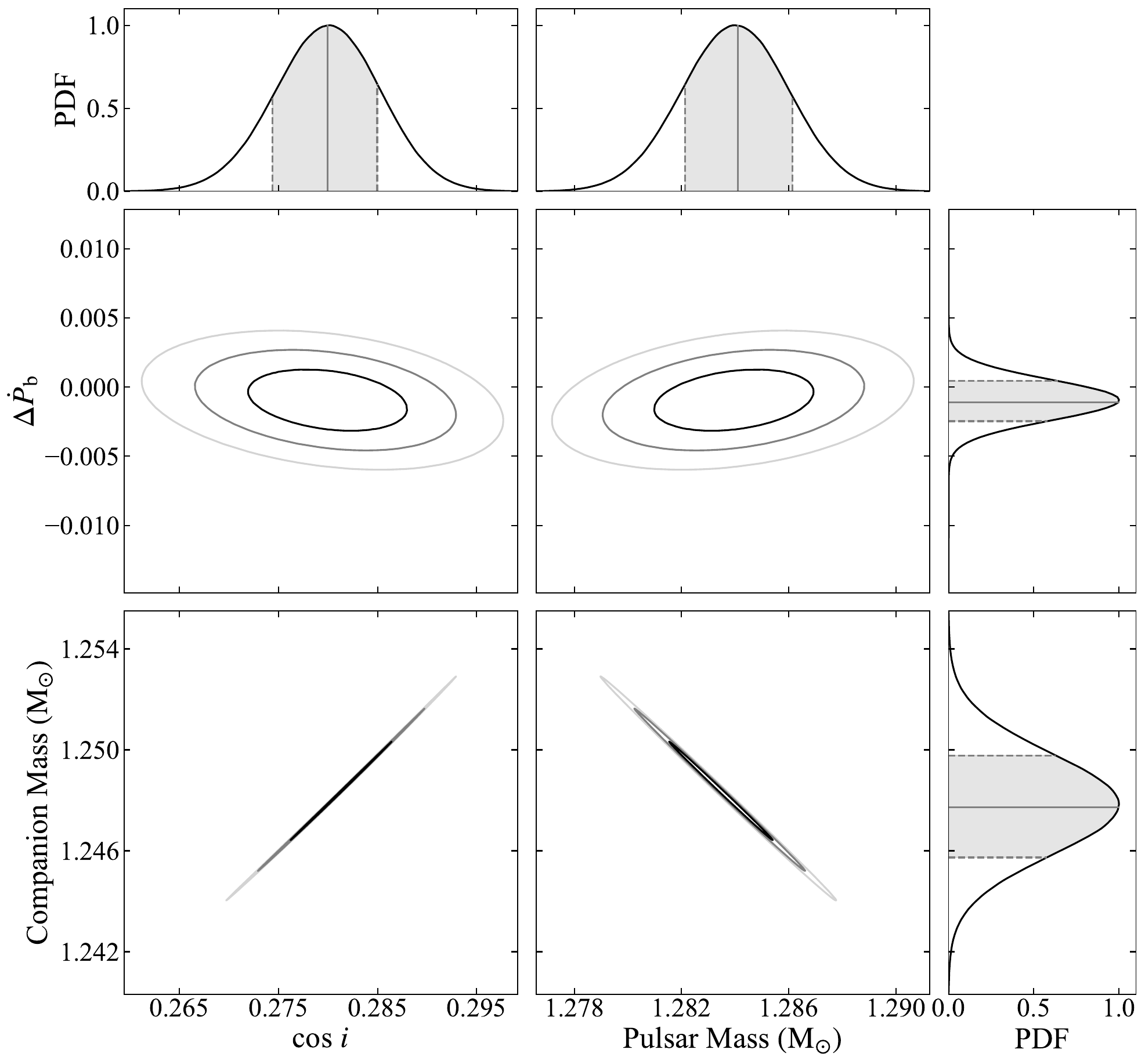}
    \caption{Constraints on the orbital inclination angle, masses of the pulsar and the companion and the $\dot{P}_{\rm b}$ deviation from the measurement and the prediction of GR ($\Delta \dot{P}_{\rm b}$). The black, grey and light grey contours represent the 1-$\sigma$, 2-$\sigma$ and 3-$\sigma$ confidence levels of these parameters. We show the probability density of each parameter on the upper and right edges, where the grey area and solid grey line represent the 1-$\sigma$ confidence area and the nominal values; the dashed lines at the edges represent the $\pm 1$-$\sigma$ limits.}
    \label{fig:chisqmap}
\end{figure*}

We will now elaborate on these two aspects.

\subsection{Masses of the two neutron stars}
\label{sec:masses_bayesian}

Using the DDGR model, which assumes the validity of GR (and furthermore, that the observed PK parameters are relativistic), we can fit directly for the masses. Doing this, we obtain $M = 2.531837(24)\,\rm M_{\odot}$ and $m_{\rm c} = 1.2476(20)\,\rm M_{\odot}$, thus $m_{\rm p} = 1.2842(21)\,\rm M_{\odot}$. Unsurprisingly, these 
values coincide exactly with our estimates above based on $\dot{\omega}$ and $\gamma$; this indicates that it is the presence of the relativistic effects associated with $\dot{\omega}$ and $\gamma$ in the timing of the pulsar that is chiefly responsible for the precise DDGR mass estimates.
It corresponds to $\sin i = 0.960(2)$ and the orbital inclination angle $i = 73.7(3)^{\circ}$ (or $106.3(3)^\circ$).

In order to better estimate the uncertainties in these parameters, we made a $\chi^2$ map in a
3-dimensional space where the coordinates are $M$, $\cos i$ and $\Delta\dot{P}_{\rm b}$. This used the same basic Bayesian methodology as \cite{2002ApJ...581..509S}, where the $\chi^2$ obtained by a DDGR model at each point of that space (with the corresponding values of MTOT, M2 and XPBDOT fixed to their values at that point)
is used to derive a 3-D probability distribution function (pdf). In Fig.~\ref{fig:chisqmap}, this is projected into four 2-D spaces that have
$\cos i$, $\Delta\dot{P}_{\rm b}$, $m_{\rm c}$ and $m_{\rm c}$ as axes, and finally the pdfs are marginalised along those
axes. From this analysis, we obtain: $M = {2.53186(4)\,\rm M_{\odot}}$, $m_{\rm c} = {1.248(2)\,\rm M_{\odot}}$, $m_{\rm p} = {1.284(2)\,\rm M_{\odot}}$, $\cos i = 0.280(5)$ and $\Delta\dot{P}_{\rm b} = -0.9^{+1.4}_{-1.5} \,\rm fs\,s^{-1}$. These estimates and their uncertainties agree with the simpler estimates above, indicating that the uncertainties obtained by the simple DDGR fit are essentially accurate.

Apart from measuring the masses and their uncertainties, the detailed mapping of the allowed masses allows a precise prediction of $\dot{P}_{\rm b, GR}$ and an estimate of $\Delta \dot{P}_{\rm b}$,
where the correlation between $m_{\rm c}$ and $m_{\rm p}$ is taken into account.

\subsection{Testing GR with the orbital decay}

The masses we used in the previous section are calculated from $\dot{\omega}$ and $\gamma$ that are derived from the DDFWHE model.
Then, we use these masses to calculate the predicted orbital decay under GR ($\dot{P}_{\rm b, GR}$).
The difference between $\dot{P}_{\rm b,int}$ and $\dot{P}_{\rm b, GR}$ ($\dot{P}_{\rm b, xs}$) is at most $0.3 \pm 1.6 \rm \, fs\,s^{-1}$, which indicates a non-detectable difference. The ratio of $\dot{P_{\rm b}}^{\rm int}$ to $\dot{P}_{\rm b, GR}$ is given by:
\begin{equation}
    \frac{\dot{P}_{\rm b, int}}{\dot{P}_{\rm b, GR}}=\frac{-1.8288(16)\times10^{-12}}{-1.8287(1)\times10^{-12}}=1.00005(91),
\end{equation}
thus $\dot{P}_{\rm b,int}$ is fully consistent with the predicted value, $\dot{P}_{\rm b, GR}$.

This radiative test is about twice as precise as the test with the Hulse-Taylor pulsar \citep{2016ApJ...829...55W,2018ApJ...862..139D}, making it the second most precise test of the quadrupole formula; however, it is still one order of magnitude behind the precision achieved with the Double Pulsar \citep{2021PhRvX..11d1050K}.

\begin{figure}
    \centering
    \includegraphics[width=\linewidth]{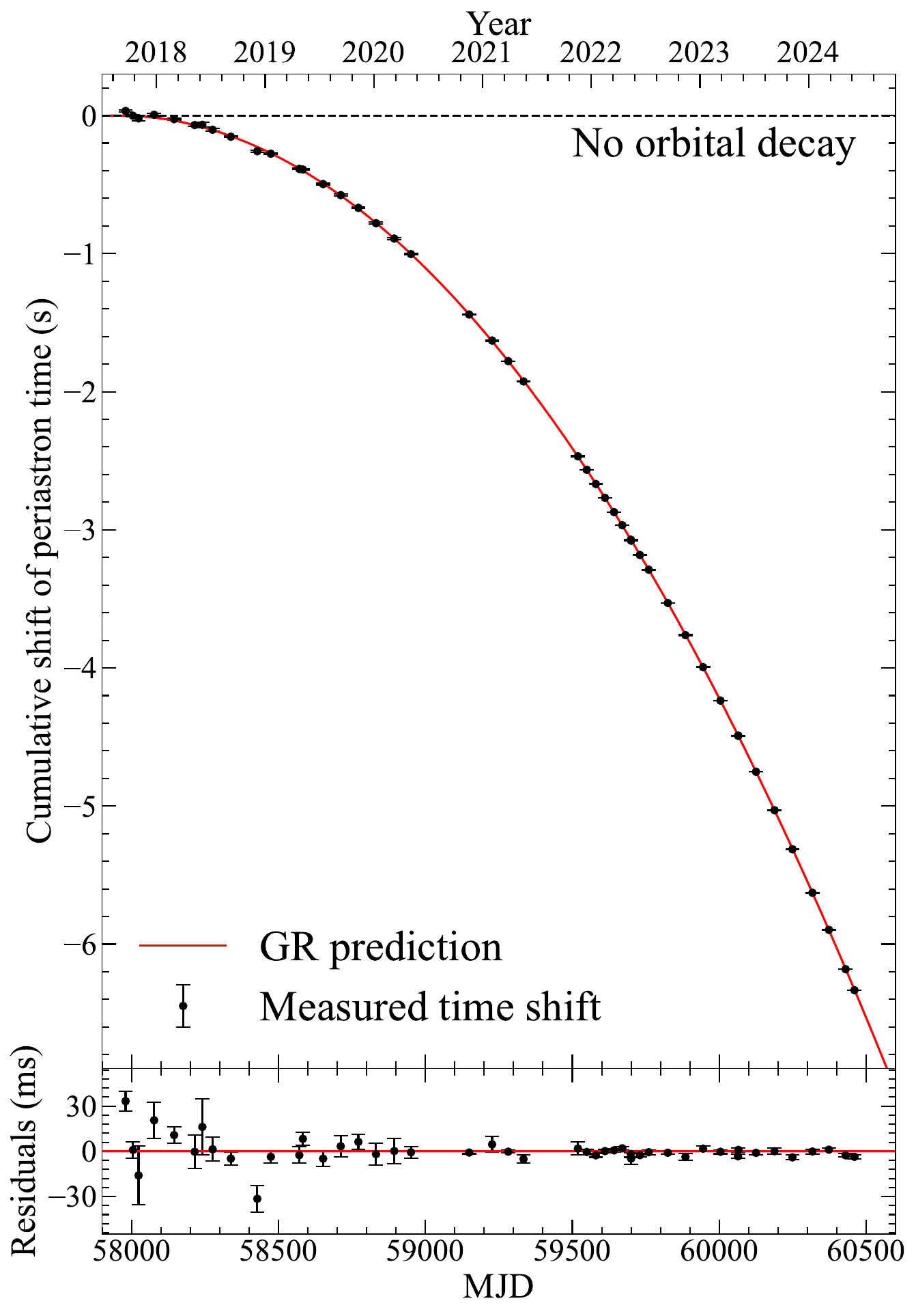}
    \caption{The cumulative shift of periastron time relative to the predictions of a constant orbit caused by the decay of the orbital period. In the upper panel, the black points are the measurements of the periastron time shift, and the red solid line is the GR prediction, which is the sum of the GW damping calculated from GR and the calculated contribution from the Galactic acceleration and the Shklovskii effect. We display the residual between the measurements and the GR prediction in the lower panel.}
    \label{fig:orbital decay}
\end{figure}

In Fig.~\ref{fig:orbital decay}, we see the cumulative shift of periastron time. This is the difference between the time of passage through the periastron measured at any particular time and what its value would be if the orbital period had stayed constant at its value at the start of the timing. The linear change in the orbital period predicted by GR results in a quadratic shift in the time of periastron that is indicated by the red solid line. In black, we measured the times of passage of periastron with subsets of
TOAs, where all other timing parameters are fixed. As we can see, there is an excellent match between the GR prediction and the measurements of $T_0$. This plot does not only tell us that the magnitude of the orbital decay is as expected by GR, but also shows that this decay proceeds linearly with time as expected from a constant rate of loss of orbital energy.

Note that the result is derived from 7 years of observations with Arecibo, GBT and FAST. 
Considering the observation duration of 31 years for the Hulse-Taylor pulsar and 16 years for the Double Pulsar, with the follow-up observations of FAST, PSR~J1946+2052 has the potential to increase the precision of its test significantly as the measurement for $\dot{P}_{\rm obs}$ improves according to $T_{\rm obs}^{5/2}$ where $T_{\rm obs}$ is the observed time span \citep{1992PhRvD..45.1840D}.

\subsection{Testing GR with the Shapiro delay}

The $\dot\omega$-$\gamma$-$\dot{P}_\mathrm{b}$ test of the previous section is a mixed test which combines quasi-stationary and radiative strong field effects. The additional detection of the Shapiro delay in the PSR~J1946$+$2052 system allows for a different type of test, a purely quasi-stationary test that combines orbital and signal propagation effects. Therefore, such a test probes different aspects of gravity compared to the $\dot\omega$-$\gamma$-$\dot{P}_\mathrm{b}$ test, in particular the propagation of electromagnetic signals in the spacetime of a strongly self-gravitating body (cf.\ \citealt{1992Natur.355..132T, Damour_2009, Wex_2014}). Although the measurements of the Shapiro delay parameters $h_3$ and $\varsigma$ are much less significant than the measurement of $\dot{P}_{\rm b}$, they still provide two tests for GR with good precision. The GR predictions are calculated from Eq.~(\ref{eq:gamma_masses}), (\ref{eq:sini}) and (\ref{eq:shapiro}), where $M$ and $m_{\rm c}$ were derived from $\dot{\omega}$ and $\gamma$ using the GR equations. Dividing the measured parameters by their GR predictions, we obtain:
\begin{align}
    &\frac{h_3}{h_{3, \rm GR}}=\frac{2.58(14)\times10^{-6}}{2.59(4)\times10^{-6}}=0.997(56),
    \\&\frac{\varsigma}{\varsigma_{\rm GR}}=\frac{0.760(36)}{0.750(4)}=1.01(5),
\end{align}
indicating a good consistency.

We can compare these with the GR tests provided by the $r$,$s$ parametrisation of the DD model:
\begin{align}
    &\frac{m_{\rm c, S}}{m_{\rm c, GR}} =\frac{1.21(21)\,{\rm M_\odot}}{1.248(2)\,{\rm M_\odot}}=0.97(15),
    \\&\frac{\sin i_{\rm S}}{\sin i_{\rm GR}} =\frac{0.962(13)}{0.9599(16)}=1.002(13).
\end{align}
The results are more precise than those in the Hulse-Taylor pulsar \citep{2016ApJ...829...55W} but less precise than PSR~B1534+12 \citep{2014ApJ...787...82F} and the Double Pulsar \citep{2021PhRvX..11d1050K}, owing to the intermediate orbital inclination angle of 73.7(3)$^\circ$.

These measurements provide a direct comparison of the $h_3$ and $r$ tests, with the former being 2.5 times more precise, apart from being less correlated with $\varsigma$ than $r$ is to $s$. This illustrates the fact highlighted by \cite{2010MNRAS.409..199F} that, by decreasing the correlation between the parameters that quantify the Shapiro delay, the orthometric parametrisation provides an improved GR test relative to the ``classical'' $r$-$s$ parametrisation. The difference between these two tests can be illustrated in Fig.~\ref{fig:mmdiagram}: in this diagram, the $r$ constraint from Shapiro delay ($m_{\rm c} = 1.21(21)\,{\rm M_\odot}$) would occupy a much wider region of the diagram compared to the $h_3$ band.

Note, on the other hand, that the $s$ test seems to be more precise than the $\varsigma$ test. This is a purely numerical artefact, caused by the fact that as angles come closer to $90^\circ$, the variations of $\sin i$ become very small. As discussed in Sect.~\ref{subsec:shapiro}, the $\varsigma$ and $s$ constraints result in very similar orbital inclinations, which implies similar bands for these two parameters in Fig.~\ref{fig:mmdiagram}.

\begin{figure}
    \centering
    \includegraphics[width=\linewidth]{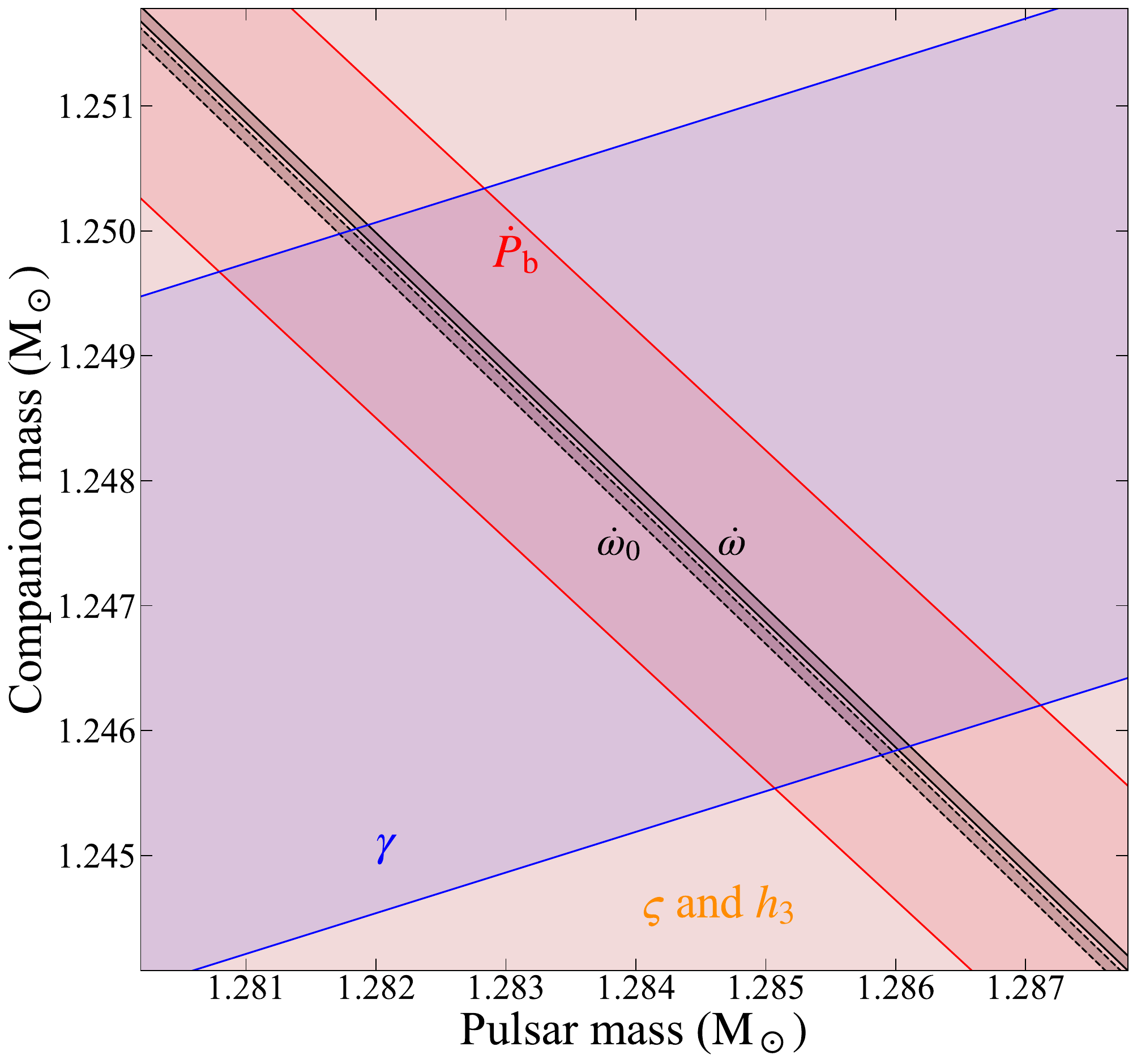}
    \caption{The enlarged mass-mass diagram for PSR~J1946+2052. To demonstrate the influence introduced by the LT effect, we display the total mass constraint for $\dot{\omega}_0$ with black dashed lines, where we ignore the LT contribution.
    While the black solid lines represent $\dot{\omega}$ with the full contribution, including the 2PN and LT effect.
    The significant deviation between $\dot{\omega}_0$ and $\dot{\omega}$ indicates that the LT contribution significantly affects our measurement of the total mass. The magnitude of this shift
    depends on the MOI of the pulsar.
    Owing to the much larger uncertainty of $\dot{P}_{\rm b}$, the total mass cannot be measured independently with sufficient precision to determine the MOI.}
    \label{fig:mmdiagram_2}
\end{figure}

\subsection{Effect of higher-order contributions on the measurement of the total mass}
\label{sec:omdot_NLO}

Given the precise measurement of $\dot{\omega}$, we can explore the higher-order corrections on this parameter.
The $\dot{\omega}$ can be written by:
\begin{equation}
    \dot{\omega}=\dot{\omega}^{\rm 1PN} + 
    \dot{\omega}^{\rm 2PN} + \dot{\omega}^{\rm LT},
    \label{eq:om}
\end{equation}
where $\dot{\omega}^{\rm 1PN}$ and $\dot{\omega}^{\rm 2PN}$ are the first and second post-Newtonian (PN) contributions, ignoring spin contributions. 
$\dot{\omega}^{\rm LT}$ is the Lense-Thirring (LT) precession contribution caused by the coupling between the spins of the binary components and the orbital motion.
Note that the LT term in Eq.~(\ref{eq:om}) is completely dominated by the spin of the (observed) pulsar due to the expected large spin period of the companion (see details in Sect.~\ref{sec:pbdot}).

The 1PN and 2PN contributions of $\dot{\omega}$ are given by \citep{Robertson_1938,damour1985general,damour1986general,Damour_1987,1988NCimB.101..127D}:
\begin{align}
    \label{eq:omdot1pn}
    &\dot{\omega}^{\rm 1PN}=3n^{\frac{5}{3}}_{\rm b}(T_{\odot}M)^{\frac{2}{3}}(1-e^2)^{-1},
    \\& \dot{\omega}^{\rm 2PN}=3n^{\frac{7}{3}}_{\rm b}(T_{\odot}M)^{\frac{4}{3}}(1-e^2)^{-1}f_{\rm O},
\end{align}
and 
\begin{equation}
\begin{aligned}
    \label{eq:fo}
    f_{\rm O}=
    &\frac{1}{1-e^2}\left(\frac{39}{4}X_{\rm p}^2+\frac{27}{4}X_{\rm c}^2+15X_{\rm p}X_{\rm c} \right)
    \\& -\left(\frac{13}{4}X_{\rm p}^2+\frac{1}{4}X_{\rm c}^2+\frac{13}{3}X_{\rm p}X_{\rm c} \right),
\end{aligned}
\end{equation}
where $X_{\rm p} \equiv {m_{\rm p}}/M$ and $X_{\rm c} \equiv {m_{\rm c}}/M$.

If the misalignment angle between the spin axis and the orbital angular momentum is small
(in the case of PSR~J1946+2052 \citealt{2024ApJ...966...46M}
estimate a misalignment angle of only $\sim 0.2 \deg$) then $\dot{\omega}^{\rm LT}$ is given by \citep{1975PhRvD..12..329B,1988NCimB.101..127D,2024LRR....27....5F}:
\begin{equation}
    \label{eq:omdotlt}
    \dot{\omega}^{\rm LT}=
    -n_{\rm b}^{2}(T_{\odot}M)(1-e^2)^{-3/2} (3+X_{\rm p})X_{\rm p}\,\chi_{\rm p},
\end{equation}
where $\chi_{\rm p}$ is the dimensionless spin of the pulsar, given by
\begin{align}
    \label{eq:chia}
    \chi_{\rm p}=\frac{G}{c^5}(T_{\odot}m_{\rm p})^{-2} \, 2\pi\nu_{\rm p}I_{\rm p},
\end{align}
and $\nu_{p}$ is the spin frequency of the pulsar.

From these equations, and the masses of the components of the system, we derive $\dot{\omega}^{\rm 2PN} = 0.00078422(2)\,{\rm deg\,yr^{-1}}$ and $\dot{\omega}^{\rm LT} = -0.000884(1)\times I_{\rm p}^{(45)}\,{\rm deg\,yr^{-1}}$.
One thing that is important to emphasise is that both values are significantly larger than the corresponding values for PSR~J0737$-$3039A. In particular, for the latter pulsar, \cite{2021PhRvX..11d1050K} estimate
$\dot{\omega}^{\rm LT} = -0.000377\times I_{\rm A}^{(45)}\,{\rm deg\,yr^{-1}}$,
a factor of 2.3 smaller than for PSR~J1946+2052, which can only be in a small part compensated by the slightly larger
value of $I_{\rm A}$ compared to $I_{\rm p}$.
The reason is that, as shown by Eq.~(\ref{eq:omdotlt}), the effect increases with the square of the
orbital frequency, and thus (according to Kepler's law) with the inverse cube of the orbital separation of the masses. Furthermore, as shown by Eq.~(\ref{eq:chia}), the spin frequency is important, and PSR~J1946+2052 has the fastest spin of any known pulsar in a DNS system.

A consequence of the magnitude of these effects is that they are already larger than the uncertainty of $\dot{\omega}$, $\delta \dot{\omega} = 0.00040\,{\rm deg\,yr^{-1}}$. This implies that to measure $M$ and its uncertainty accurately (in GR), we have to take the effect of $\dot{\omega}^{\rm 2PN}$ and $\dot{\omega}^{\rm LT}$ into account, as in the case of the PSR~J0737$-$3039 system \citep{2021PhRvX..11d1050K}.
The value of $I_{\rm p}^{(45)}$ is determined as 1.31(8) by employing different equations of state (see Appendix \ref{sec:appendixA} for details), indicating that $\dot{\omega}^{\rm LT} = -0.00117(7)\,{\rm deg\,yr^{-1}}$, which is ~3 times larger than $\delta\dot{\omega}$.

The total mass of the system in Sects.~\ref{sec:results} and \ref{sec:masses} are derived from $\dot{\omega}$ with only 1PN contribution. 
Now we start to consider the 2PN and LT contributions.
One way to distinguish the significance of the LT contribution in $\dot{\omega}$ is to ignore the LT term and assume the observed $\dot{\omega}$ consists of 1PN and 2PN contributions.
By doing this, we can plot the curve of $\dot{\omega}$ in the mass-mass diagram, as indicated by $\dot{\omega}_0$.
Then we plot the observed $\dot{\omega}$ with the full expression, including $\dot{\omega}^{\rm 2PN}$ and $\dot{\omega}^{\rm LT}$.
From Fig.~\ref{fig:mmdiagram_2}, one notes that the 1$-\sigma$ areas of $\dot{\omega}_0$ and $\dot{\omega}$ are already separated, which indicates different values of the total mass of the system when we consider the LT contribution or not.
In Table~\ref{tab:omdot}, we present the calculations of the masses with different contributions in $\dot{\omega}$.
The masses of the pulsar and the companion are all consistent at 1$-\sigma$ confidence level, indicating that the analyses in the previous sections will not be affected by the higher-order contributions of $\dot{\omega}$.
However, the total mass shows a significant difference, which means the 2PN and LT contributions have already had impact on the measurement of the total mass.

\begin{table}[]
    \centering
        \caption{The total mass of PSR~J1946+2052, the pulsar mass and the companion mass considering different orders of contribution in $\dot{\omega}$.}
    \begin{tabular}{l c c c}
    \hline
    \hline
         &  $M$  &  $m_{\rm c}$     &  $m_{\rm p}$\\
         & \multicolumn{3}{c}{($\rm M_{\odot}$)} \\
    \hline 
    
    $\dot{\omega}^{\rm 1PN}$
    &   2.531858(60)    &  1.2480(21)     &  1.2838(21) \\
    $\dot{\omega}^{\rm 1PN}+\dot{\omega}^{\rm 2PN}$
    &   2.531754(60)    &  1.2478(20)     &  1.2839(20) \\
    $\dot{\omega}^{\rm 1PN}+\dot{\omega}^{\rm 2PN}+\dot{\omega}^{\rm LT}$
    &   2.531922(57)    &   1.2477(21)    &   1.2842(21) \\
    \hline
    \end{tabular}

    \label{tab:omdot}
\end{table}

Another possible source of uncertainty is the contribution from the proper motion. Rewriting the equations of
\cite{1996ApJ...467L..93K}, one obtains:
\begin{equation}
\dot{\omega}_{\rm K} = \frac{\mu}{ \sin i} \cos(\Theta_{\mu} - \Omega),
\end{equation}
where $\Theta_{\mu}$ and $\Omega$ are the position angles of the proper motion and the line of nodes, respectively.
Using the values above, we obtain at most a contribution of $1.3 \times 10^{-6}\, \deg \, \rm yr^{-1}$, which is two orders of magnitude smaller than the measurement uncertainty.

\section{Conclusions}
\label{sec:conclusions}

In this work, we have presented a comprehensive timing analysis of the DNS system PSR~J1946+2052 using the observations from Arecibo, GBT and FAST.
After dealing with the profile evolution caused by the relativistic spin precession of the pulsar, we derived ToAs from the 7-year period of observations.
These were used to derive precise timing solutions using DDFWHE and DDGR binary models.

This led to precise measurements of the proper motion and of five PK parameters, enabling a robust determination of precise masses for the pulsar and the companion and three tests of general relativity. The observed orbital decay deviates from the GR prediction by less than $9 \times 10^{-4}$ (68\% C. L.), making this the second most precise test of the quadrupole formula to date.
The Shapiro delay measurement in this pulsar produced two additional GR tests. General relativity passes these three tests.

Owing to the short orbital period, the higher-order contributions (2PN and LT effect) to $\dot{\omega}$ are significantly larger than those for the Double Pulsar system.
By adopting multiple EOSs and multi-messenger constraints on the radius of the neutron star, we derived the moment of inertia and the LT effect on $\dot{\omega}$, which is three times larger than the measurement uncertainty for $\dot{\omega}$. This means that these higher-order contributions must be taken into account to derive accurate estimates of the total mass of the system.

Future observations from FAST will continue to rapidly improve the precision of
$\dot{P}_{\rm b, obs}$, however, the lack of precise distance measurements that are independent from the Galactic electron density models will preclude a significant reduction of the uncertainty of $\dot{P}_{\rm b, ext}$, which is now of the order of 0.7 fs/s, or about half the current uncertainty of $\dot{P}_{\rm b, obs}$. Thus, to improve the precision of the radiative test (i.e., of $\dot{P}_{\rm b, ext}$ and $\dot{P}_{\rm b, int}$) by more than a factor of 2, a precise independent measurement of the distance will be necessary.
This uncertainty of the measurement of $\dot{P}_{\rm b, int}$ precludes a precise independent determination of the total mass of the system, which would be necessary to extract $\dot{\omega}^{\rm LT}$ from the total observed $\dot\omega$ and determine the MOI of the pulsar.
The future VLBI observations operated by SKA will greatly improve the measurement of the distance, and then improve the precision of $\dot{P}_{\rm b, int}$, which will then provide a more precise radiative test and possibly allow a measurement of the MOI for this pulsar.

Another method to determine the MOI is measuring the LT contribution in $\dot{x}\  (\dot{x}^{\rm LT})$, a result of the LT-induced secular variation of the orbital inclination $i$ (see e.g.\ \citealt{1988NCimB.101..127D,1992PhRvD..45.1840D}).
If future observations confirm the result in \cite{2024ApJ...966...46M}, at least in its order of magnitude, then the maximum of $\dot{x}^{\rm LT}$ is of the order of $3\times10^{-16}\,\rm s\,s^{-1}$, which is considerably smaller than the current constraint on $\dot{x}$ with a level of $3\times10^{-14}\,\rm s\,s^{-1}$.

The ongoing observations of PSR~J1946+2052 will continue monitoring the observed changes of the pulse profile and polarisation \citep{2024ApJ...966...46M}, which not only will improve the constraints on the orbital geometrical parameters, including solid measurement of the misalignment angle, but might also result in an additional test of GR in this system, via the measurement of the geodetic precession rate. 
This test might be particularly favoured since for PSR~J1946+2052 we detect a strong interpulse (see Fig.~\ref{fig:profile}); in most such cases one can achieve an improved determination of the spin geometry from the rotating vector model, which results in improved measurements of the geodetic precession rate \citep{stairs2004measurement,Desvignes_2019}.

Additionally, continued observations will
greatly improve the precision of the proper motion, which will improve the estimate of the pulsar's transverse velocity relative to that of its Galactic co-rotating frame. The combination of the spatial velocity and the misalignment angle will yield estimates of the kick associated with the second supernova in this system and provide an improved understanding of the evolution of the PSR~J1946+2052 system.

\begin{acknowledgements} 

We thank Cijie Zhang and Yukai Zhou for deploying and maintaining the server used throughout this research.
This work was supported by the National Natural Science Foundation of China (12041303, 12421003, 12203072, 12203070), the National SKA Program of China (2020SKA0120200, 2020SKA0120300), Beijing Nova Program (No. 20250484786)
 and the CAS-MPG LEGACY project. 
Pulsar research at UBC is supported by an NSERC Discovery Grant and by the Canadian Institute for Advanced Research.
LM gratefully acknowledges the support of the China Scholarship Council (CSC) during the visit to the Max-Planck Institute for Radio Astronomy in Bonn, Germany. 
PCCF gratefully acknowledges continuing support from the Max-Planck-Gesellschaft and the hospitality of the Academia Sinica Institute of Astronomy and Astrophysics in Taipei, where part of this work was conducted while he was a Visiting Scholar.
EP is supported by a Juan de la Cierva fellowship (JDC2022-049957-I).
LS was supported by the Max Planck Partner Group Program funded by the Max Planck Society.
MAM is supported by the NANOGrav Physics Frontiers Center (NSF Award Number 2020265)
JY was supported by the National Science Foundation of Xinjiang Uygur Autonomous Region (2022D01D85), the Major Science and Technology Program of Xinjiang Uygur Autonomous Region (2022A03013-2), the Tianchi Talent project, and the CAS Project for Young Scientists in Basic Research (YSBR-063), the Tianshan talents program (2023TSYCTD0013), and the Chinese Academy of Sciences (CAS) “Light of West China” Program (No. xbzg-zdsys-202410  and No. 2022-XBQNXZ-015).
This work made use of the data from FAST (Five-hundred-meter Aperture Spherical radio Telescope). FAST is a Chinese national mega-science facility, operated by National Astronomical Observatories, Chinese Academy of Sciences.
While taking data for this work, the Arecibo Observatory was operated by SRI International
under a cooperative agreement with the National Science
Foundation (NSF; AST-1100968), and in alliance with Ana G. M\'endez-Universidad Metropolitana, and the Universities Space Research Association. The National Radio Astronomy Observatory is a facility of the NSF operated under cooperative agreement by Associated Universities.
We thank the staff of the Arecibo Observatory for all the help during this project, especially Arun Venkataraman. 

\end{acknowledgements} 

\bibliography{aa55689-25}
\bibliographystyle{aa}

\begin{appendix}

\section{Observed DM variations and their power spectrum}
\label{sec:sf}

DM variation is a common phenomenon in pulsar astronomy, starting from the first detection of such epoch-dependent variations in the Crab Pulsar \citep{1971IAUS...46..114R}.
The solar wind can contribute to DM variation, which can be corrected, to first order, by \texttt{TEMPO} by 
assuming a constant electron density of the Solar wind at 1 au from the Sun; this is by default set at $10\, \rm cm^{-3}$. 
Due to the elliptical latitude of 41.25$^\circ$, the impact of the solar wind could be negligible.
The more significant contribution could be the variation of the large DM of 93.9281(9)$\,\rm pc\,cm^{-3}$ of PSR~J1946+2052.

\begin{figure}
    \centering
    \includegraphics[width=1.\linewidth]{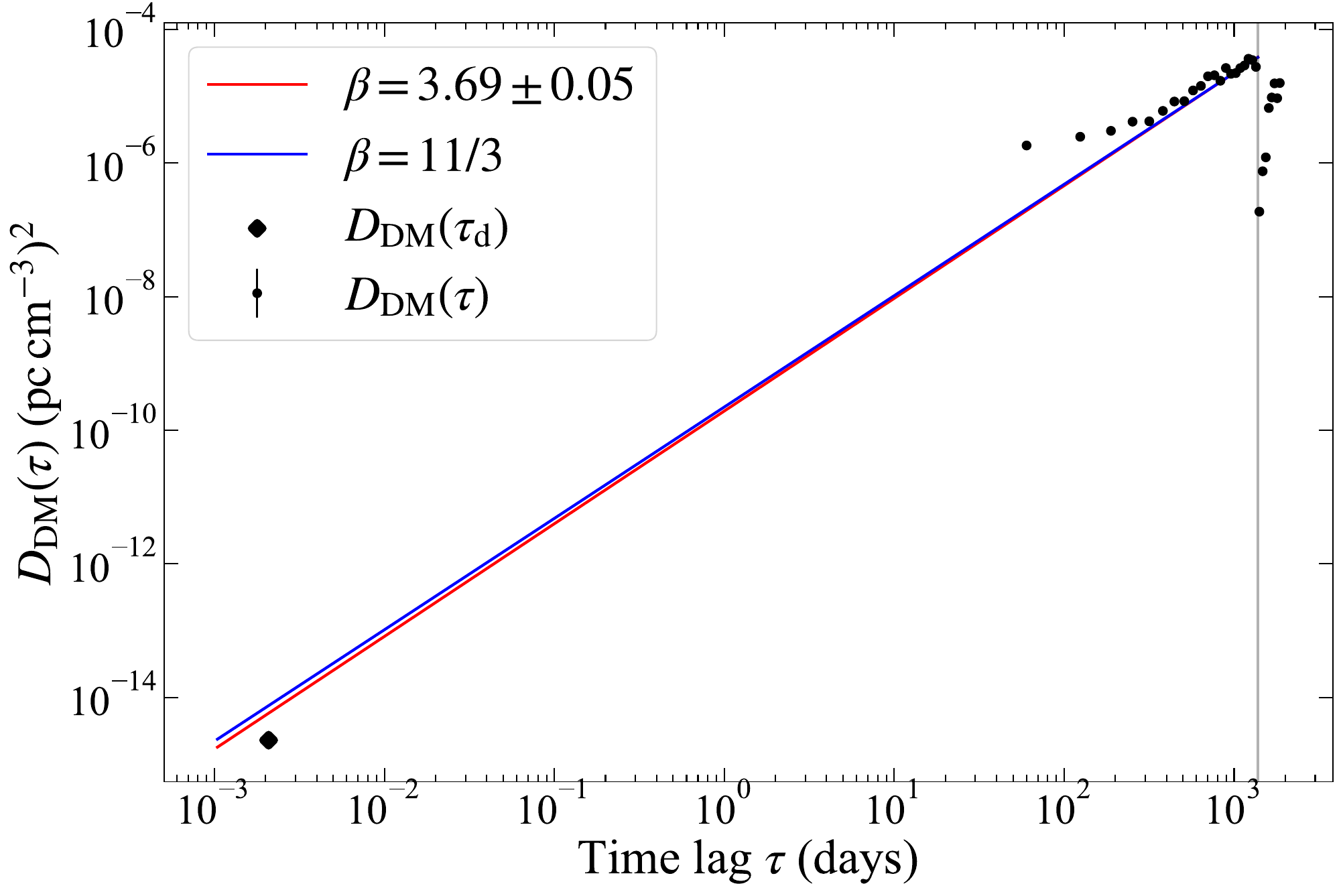}
    \caption{The measurements of DM structure functions as a function of time lag are displayed in black circles. The black diamond represents the DM structure function when the time lag equals the time scale of DISS ($\tau_{\rm d}\approx 3$ minutes), which is included in the fitting. The red and blue dashed lines are the fitting result with $D_{\phi}(\tau)\propto \tau^{\beta-2}$ and fixed $\beta$ of 11/3, respectively. Only points to the left of the grey line are involved in the fitting due to the possible bias introduced by the sudden decrease.}
    \label{fig:sf}
\end{figure}

As mentioned previously, these variations are due to the fact that the motion of the pulsar and the Earth changes the path followed by the radio waves from the pulsar to the Earth; these changing paths imply that, at different times, the radio waves go through slightly different regions of the IISM. The resulting stochastic variation of the DM is an interesting probe of the spatial structure of the IISM.

One way to characterise this stochastic variation is by using the spatial power spectrum, which is a function of power-law wavenumber\citep{1990ARA&A..28..561R}:
\begin{equation}
    P(q)=C_{n}^{2}q^{-\beta};\ \ \ \ 2\pi/l_{\rm o} < q < 2\pi/l_{\rm i},
\end{equation}
where $C_{n}^{2}$ is the scaling factor of the power spectrum, $\beta$ is the power-law index, $l_{\rm o}$ and $l_{\rm i}$ are the outer and inner scales of the turbulence.
This equation assumes that the electron-density fluctuations within the interstellar medium are isotropic and the magnitude of the wavenumber determines the spectrum \citep{2016ApJ...821...66L}.
Then the power spectrum can be estimated by the phase structure function (SF) with 
\begin{equation}
    D_{\phi}(\tau)=\langle[{\phi}(t+\tau)-{\phi}(t)^2]\rangle,
\end{equation}
where $\phi(t)$ is the geometrical phase delay that indicates the spatial variations in the electron density along the line of sight between the pulsar and the observer in a specific time, $\tau$ is the time lag between phases, and the angle brackets denote an ensemble average.
If the spectrum index satisfies $2<\beta<4$ and the spatial scale is between outer and inner of the IISM that governs the power spectrum of electron density fluctuations, the phase SF can be given by $D_{\phi}(\tau)= (\tau/\tau_{\rm d})^{\beta-2}$ \citep{1977ARA&A..15..479R}, where $\tau_{\rm d}$ is the time scale of the diffractive interstellar scintillation (DISS).
Based on this, we can also define DM SF to describe DM variations:
\begin{equation}
    D_{\rm DM}(\tau)=\langle[{\rm DM}(t+\tau)-{\rm DM}(t)^2]\rangle,
\end{equation}
where the $\tau$ here is the time lag between the DM measurements.
The DM SF can be related to the phase SF with 
\begin{equation}
    D_{\phi}(\tau)=\left( \frac{2{\pi}C}{f}\right)^2D_{\rm DM}(\tau),
\end{equation}
where $C=4.148808\times10^{3} \,{\rm MHz^2\,pc^{-1}\,cm^3}$ and $f$ is the frequency of the observation in MHz.

The observed DM variations are shown in Fig.~\ref{fig:DM_variation}.
Given the systematic DM offset between DM measurements from Arecibo and FAST, we only use FAST measurements to analyse the DM variation. 
We display the result of the DM SF in Fig.~\ref{fig:sf}, where we included in the fit the DM SF when the time lag equals the time scale of DISS.
This timescale ($\tau_{\rm d}$) is estimated to be $\sim3$ minutes by using the dynamic spectrum of the observation with the highest signal-to-noise ratio.
The value of the DM SF for $\tau=\tau_{\rm d}$ is then calculated with $D_{\phi}(\tau_{\rm d})=1$.
In addition, Fig.~\ref{fig:sf} shows the power-law fit to the DM SF. The resulting spectral index of the SF is 3.69 $\pm$ 0.05, which is consistent with the typical Kolmogorov spectrum with an index of 11/3 ($\sim3.667$), indicating that the interstellar medium between the Earth and the pulsar is incompressible, homogeneous and isotropic.

\section{The moment of inertia of the pulsar}
\label{sec:appendixA}

To calculate the Lense-Thirring contribution to the periastron advance caused by the fast-rotating pulsar, one needs to know the MOI of PSR~J1946+2052, $I_\mathrm{p}$. Given the rather precise value for the pulsar mass $m_\mathrm{p}$ (see Tab.~\ref{tab:timsol}), the MOI is known with high precision once an EOS has been chosen to describe the density-pressure relation for neutron star matter. Unfortunately, our knowledge of the properties of matter at supranuclear densities is still limited, leading to a considerable uncertainty in our knowledge of the true EOS for neutron stars. This is reflected in a corresponding inaccuracy in our knowledge of the MOI of a neutron star with given mass. The purpose of this section is to obtain a probability distribution for $I_\mathrm{p}$ that takes into account this imperfect knowledge of the density structure of PSR~J1946+2052.

\begin{figure}
    \centering
    \includegraphics[width=\linewidth]{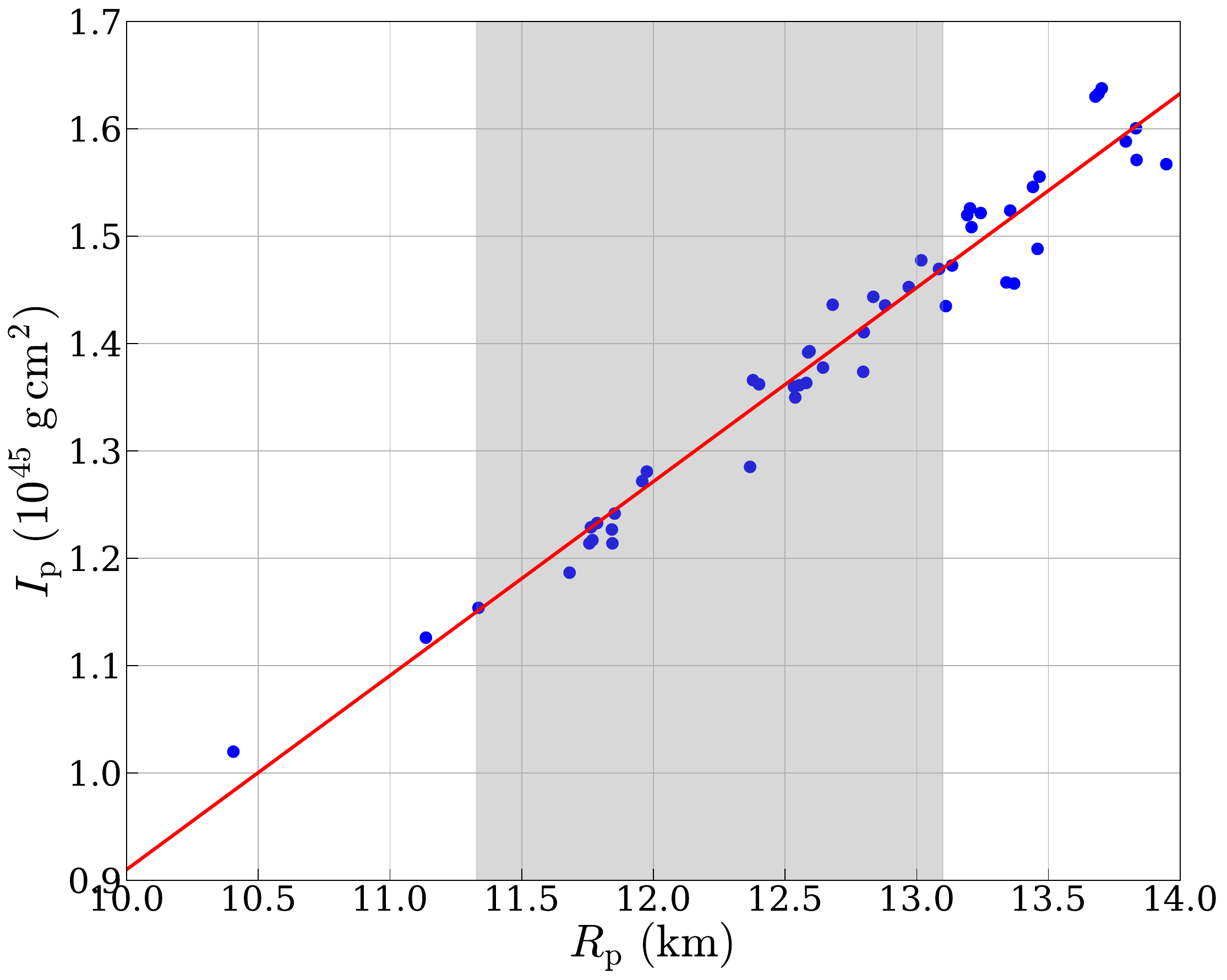}
    \caption{Moment of inertia of PSR~J1946+2052 (mass = 1.284\,$\Msun$) as a function of the pulsar's radius, for 50 EOSs, which all have a maximum neutron-star mass exceeding $1.96~\Msun$. The red line shows the best linear fit to the data points. The grey range indicates the conservative 95\% credibility interval of $R_{1.4}$ given by \cite{Koehn_2024}. See text for more details.}
    \label{fig:R-I45}
\end{figure}

In a first step, we use the piecewise polytropic approach of \cite{Read_2009} and calculate the MOI of PSR~J1946+2052 for all EOSs in Table~III of \cite{Read_2009} and Table~V of \cite{Kumar_2019} that have a maximum mass of at least 1.96~$\Msun$. A maximum mass below that value is excluded with 95\% confidence by the mass measurement of \cite{Fonseca_2021}.\footnote{The somewhat more model-dependent neutron star mass for PSR~J0952$-$0607 found by \cite{Romani_2022} excludes maximum neutron-star masses below 1.96~$\Msun$ with even about 99\% confidence.} 
Figure~\ref{fig:R-I45} shows this result for neutron stars with radii in the range of 10\,km to 14\,km, where $I_\mathrm{p}$ can be approximated with sufficient precision as a linear function of the radius of PSR~J1946+2052 $R_\mathrm{p}$. Note that on both ends the $R_\mathrm{p}$-range of Fig.~\ref{fig:R-I45} stretches well beyond the range of EOSs allowed by a combination of a large set of constraints by nuclear physics and multi-messenger astrophysics, as given, e.g., by \cite{Koehn_2024}.

An important quantity, when presenting constraints on EOSs is the radius of a ``canonical'' neutron star of $1.4~\Msun$. \cite{Koehn_2024} find $R_{1.4} = 12.27_{-0.94}^{+0.83}~\mathrm{km}$ with 95\% credibility as their conservative result. To a good approximation, we can adopt these limits for PSR~J1946+2052, since for most EOSs capable of supporting a $\sim 2~\Msun$ neutron star the radius changes very little when the neutron star mass is lowered from $1.4\,\Msun$ to $1.284\,\Msun$ (see e.g.\ Fig.~2 in \cite{Lattimer_2001}). For simplicity, we now assume a Gaussian probability distribution for $R_\mathrm{p}$ with the same 2-$\sigma$ bounds as $R_{1.4}$ in \cite{Koehn_2024}. Using the linear relation between $R_\mathrm{p}$ and $I_\mathrm{p}$ of Fig.~\ref{fig:R-I45} (red line), this probability distribution can be converted to a Gaussian probability distribution for the MOI of PSR~J1946+2052, which is (1-$\sigma$ error)
\begin{equation}
    I_\mathrm{p} = (1.31 \pm 0.08) \times 10^{45}~\mathrm{g\,cm^2} \,.
\end{equation}

The $\sim 6$\% uncertainty for the MOI of the pulsar leads to the same fractional error for the calculated $\dot{\omega}_\mathrm{LT}$. Given that $\dot\omega_\mathrm{LT}$ is only three times the 1$-\sigma$ error of the observed periastron advance ($\delta\dot\omega$), this error is practically negligible for the considerations in the main text, in particular Sec.~\ref{sec:omdot_NLO}. In retrospect, this also justifies some of the idealised approaches in this section.

\end{appendix}

\end{document}